\documentclass[12pt,preprint]{aastex}

\shortauthors{Tregillis{\it et al.~}}
\shorttitle{Shock Acceleration and Aging in 3D Flows}

\begin{document}

\newcommand\etal{{\it et al.~}}
\newcommand\cf{{\it cf.~}}
\newcommand\eg{{\it e.g.,~}}
\newcommand\ie{{\it i.e.,~}}

\title{Simulating Electron Transport and Synchrotron Emission in Radio 
Galaxies: Shock Acceleration and Synchrotron Aging in Three-Dimensional Flows}

\author{  I. L. Tregillis       \altaffilmark{1},
	  T. W. Jones		\altaffilmark{1},
          Dongsu Ryu            \altaffilmark{2}} 
\altaffiltext{1}{School of Physics and Astronomy, University of Minnesota,
    Minneapolis, MN 55455: tregilli@msi.umn.edu; twj@msi.umn.edu}
\altaffiltext{2}{Department of Astronomy \& Space Science, Chungnam National
    University, Daejeon, 305-764 Korea: ryu@canopus.chungnam.ac.kr}

\begin{abstract}
We present the first three-dimensional MHD radio galaxy simulations that
explicitly model transport of relativistic electrons, including diffusive 
acceleration at shocks as well as radiative and adiabatic cooling in smooth 
flows.  We discuss here three simulations of light Mach 8 jets, designed 
to explore the effects of shock acceleration and radiative aging on the 
nonthermal particle populations that give rise to synchrotron and
inverse-Compton radiations.  
Because our goal is to explore the connection between the large-scale flow 
dynamics and the small-scale physics underlying the observed emissions from 
real radio galaxies, we combine the magnetic field and relativistic electron 
momentum distribution information to compute an approximate but 
self-consistent synchrotron emissivity and produce detailed synthetic radio 
telescope observations.  

We have gained several key insights from this approach:
1. The jet head in these multidimensional simulations is 
an extremely complex environment.  The classical jet termination shock is
often absent, but motions of the jet terminus spin a ``shock-web complex'' 
within the backflowing jet material of the head.  2.  Correct interpretation 
of the spectral distribution of energetic electrons in these simulations 
relies partly upon understanding the shock-web complex, for it can give rise 
to distributions that confound interpretation in terms of the standard model 
for radiative aging of radio galaxies.  3. The magnetic 
field outside of the jet itself becomes very intermittent and filamentary in
these simulations, yet adiabatic expansion causes most of the cocoon volume 
to be occupied by field strengths considerably diminished below the nominal 
jet value.  Radiative aging is very slow in these volumes, so population
aging rates vary considerably from point to point.
4. Overall, the intricate dynamical behaviors in these models make it 
difficult to capture the histories of the nonthermal particles in broad
generalizations.  Understanding even the simplest of these models requires
attention to details of the flow.
\end{abstract}


\section{INTRODUCTION}
	
	The jet-based dynamical model for the structure of radio galaxies has 
found great support over nearly three decades of ever-increasing observational
and theoretical scrutiny.  Unfortunately, however, some of the basic physics 
attendant upon this model remains elusive.  The key idea, that lobes of 
luminous material identify the interaction between an ambient environment 
and a high-velocity plasma jet launched within an active galactic nucleus 
\citep*[\eg][]{Blandford74,Scheuer74,Begelman84,Bridle92}, accounts for  an 
impressive array of observed properties.  It is similarly well-known that the 
characteristic radio emission signals the presence of magnetic fields and 
relativistic electrons.  However, the source of these two ingredients, 
whether they are transported along the jet, or introduced in the jet-IGM 
encounter, or both, is not well understood. 
The physics of powerful radio galaxies requires not only the 
presence of relativistic particles, but also their reacceleration.  At
least in some sources, hotspot brightnesses cannot be accounted for simply
through adiabatic compression (\eg Cygnus A, Pictor A). 
Similarly, the energetics
of X-ray emission observed from hotspots and knots in some radio galaxies
also requires relativistic particle acceleration.  Yet the details of the
reacceleration process inside radio galaxies are largely unknown. 
Indeed, there is a significant 
gap in our understanding of the detailed manner in which the macrophysics of 
large-scale flows feeds the microphysics of synchrotron and X-ray emission.  

	Great advances in computational modeling have made it possible to 
perform sophisticated multidimensional hydrodynamical (HD) and 
magnetohydrodynamical (MHD) numerical simulations 
\citep*[\eg][]{Clarke89,Lind89,Kossl90,Clarke96,Nishikawa98}.
\citet*[hereafter JRE99]{Jones99} presented the first such 
simulations to include explicit time-dependent transport of the relativistic 
electrons responsible for the radio emission observed from such sources. That 
work introduced a novel, computationally economical scheme for treating 
relativistic particle transport, and applied it to ideal, axisymmetric MHD 
flows.  The scheme includes the effects of diffusive shock acceleration, 
radiative cooling, and adiabatic effects on the electrons in smooth flows.  
The addition of this new energetic particle transport feature to 
computational models of radio galaxy evolution provided a key ingredient 
needed to bridge the gap between simulation of large-scale plasma flows and 
simulation of the emissions that result.  JRE99 demonstrated that information 
about the nonthermal particle distribution could be combined with the 
magnetic field structure in the MHD flows to compute an approximate 
synchrotron emissivity in each computational zone.  
By doing so those authors were able to  
investigate the distribution of synchrotron emissivities and 
spectral indices throughout the simulated flows.  The effort of JRE99 was 
concentrated on addressing the individual dynamical influences on the shock 
acceleration of radio galaxy electrons as a way to characterize and recognize 
signs of shock acceleration in these environments.

	This paper presents the extension of that work into three dimensions.
We continue to place our emphasis upon recognizing and characterizing the 
signs of shock acceleration and radiative aging in the complex flows that are
expected to reside within radio galaxies.   This work is, therefore, a first 
attempt at combining an understanding of the flow dynamics with the subsequent
particle transport effects in three dimensions.  The notorious complexity of 
driven, multidimensional plasma flows \citep{Sato96} forces us here to restrict
our attention to a small region of the available parameter space; we defer to 
later papers the treatment of more detailed and specific questions.  Our 
long-term goal is to identify specific physical processes that lead to 
observed structures in radio galaxies, and vice-versa.

	Using the previous results in two dimensions as a guide for 
exploration, we also utilize the nonthermal particle transport scheme in a 
new way by creating the first synthetic observations to
include explicitly calculated nonthermal electron distributions.  
Preliminary results of this synthetic observation work were reported in 
\citet*[][in press]{Tregillis01a} and \citet*[][in press]{Tregillis01}. 
The earliest attempts to obtain emission characteristics from purely 
hydrodynamical calculations estimated the local magnetic field strength and
the distribution of nonthermal particles by 
assuming a simple relation with the local gas density \citep*{Smith85}.  
\citet*{Matthews90,Matthews90a} showed that the details of radio hotspot 
structures are not revealed by investigation of the interior pressure 
distribution.  However  the hotspot structures seen in images created by 
projecting the maximum density or pressure along each line of sight through 
a simulation of a precessing HD jet \citep*{Cox91} can 
show impressive similarity to the multiple hotspots observed in numerous 
real sources \citep{Lonsdale86,Looney00}. \citet*{Hughes96} concluded that 
purely HD simulations can never truly model the jet synchrotron 
emission, at least for relativistic flows.  The  inclusion of a 
vector magnetic field improves the situation by making possible the 
calculation of polarization properties \citep[\eg][]{Laing81,Clarke89}
within an assumed emissivity model.  But even fully three-dimensional MHD 
simulations require the exercise of critical assumptions before an 
emissivity can be constructed from fluid variables \citep[\eg][]{Clarke93}.
Further, even at this level of sophistication there is no means by which the 
emission spectrum can be obtained, although \citet*{Matthews90} were able to 
address the issue of deviation from a strict power law distribution of 
particles by assuming no distributed particle acceleration and introducing a 
synchrotron loss parameter associated with a fluid 
element.  The necessary information about the nonthermal particle distribution 
requires specific treatment of the essential electron microphysics that comes 
naturally out of the particle transport scheme we utilize here.

	An intermediate approach to obtaining emissivities is offered by 
Kaiser and collaborators \citep{Kaiser00a}, for example.  That work developed 
an analytical model for surface brightness distributions in the cocoons of 
FRII-type radio galaxies, based on self-similar, axisymmetric models of 
evolution.  The method extends classical spectral aging methods by taking 
into account the detailed loss histories of the emitting particles, as ours 
does explicitly, also.  Magnetic field and cosmic-ray distributions throughout
the source
must be assumed within self-similar models such as these.  In our models 
these distributions are allowed to evolve naturally from the initial 
conditions imposed at the beginning of the calculation.

	The remainder of this paper will proceed as follows.  \S 2 outlines
our methods, and provides an overview of the electron-transport scheme 
originally presented in JRE99.  The details of our models and the physical 
parameters characterizing our simulated jets are given in \S 3, and the 
results are discussed in \S 4, concentrating first upon the jet dynamics 
and then turning to consideration of the effects upon energetic particle 
transport.  The key findings are summarized in \S 5.

\section{METHODS} \label{meth.2.s}
\subsection{Dynamics} \label{dyn.2.1.s}

	We evolve the equations of ideal nonrelativistic magnetohydrodynamics 
(MHD) in Cartesian coordinates $(x, y, z)$.  All three components of velocity 
and magnetic field are included, so the model is truly three-dimensional.  The 
code is an MHD extension of Harten's \citep{Harten83} conservative, 
second-order finite difference ``total variation diminishing'' scheme, as 
detailed in \citet{Ryu95} and \citet{Ryu95a}.  The code preserves 
$\nabla \cdot B = 0$ at each time step using an approach
similar to the upwinded constrained transport (CT) scheme \citep{Evans88}
as  described in detail by \citet*{Ryu98}.  We use a passive ``mass fraction'' 
or ``color tracer'', $C_{j}$, to distinguish material entering the grid 
through the jet orifice ($C_{j} = 1$) from ambient plasma ($C_{j} = 0$).

\subsection{Electron Transport} \label{eltran.2.2.s}

	We present the general flavor of our electron transport scheme here, 
and refer the reader to the extensive discussion of the method in JRE99.  
Energetic particle transport is treated using the conventional 
convection-diffusion equation for the momentum distribution function
\begin{equation}
{\partial f \over \partial t} =
{1 \over 3}~p~{\partial f \over \partial p}(\nabla \cdot \mathbf u)-\mathbf u \cdot 
\nabla f +
\nabla \cdot (\kappa\nabla f) +
{1\over p^{2}}{\partial \over \partial p}\left(p^{2}D{\partial f \over 
\partial p}\right)
+ {\it Q}
\label{convdiff}
\end{equation}
\citep[\eg][]{Skilling75}.  
Here $\mathbf u$ is the bulk velocity of the thermal plasma, 
{\it Q} is a source 
term representing the effects of injection and radiative losses, {\it D} and 
$\kappa$ are the momentum and spatial diffusion and coefficients, and 
$f(\mathbf x,~p,~t)$ is the isotropic part of the nonthermal electron 
distribution.  This follows spatial and momentum diffusion as well as 
spatial and momentum advection of the particles.  The momentum advection 
corresponds to energy losses and gains from processes like adiabatic 
expansion as well as synchrotron aging.

	Equation \ref{convdiff} is dominated by the convective terms 
in smooth flows, because the electron diffusion lengths are much smaller
than the dynamical lengths.  However the diffusive terms are included
because when integrated across a velocity discontinuity (equation 
\ref{convdiff} is not valid inside shocks) the first three terms on the
right-hand side account for first-order Fermi acceleration at shocks
(so-called ``diffusive shock acceleration'').  The first term accounts for
adiabatic effects in the flow.  The fourth term accounts for second-order
Fermi acceleration, although we have not included it in the simulations
presented here.

	The essence of our scheme takes advantage of the fact that there is 
a strong mismatch between the scales relevant to dynamical and diffusive 
transport processes for electrons of relevance to radio and X-ray emissions
within radio galaxies.  The lengths and times appropriate to the dynamics
are orders of magnitude larger than those for electron diffusion at energies
relevant to radio synchrotron radiation.  
(Thus the convective terms in
equation \ref{convdiff} dominate the relativistic electron transport in
smooth flows.)  
As a typical example, the gyroradius
of a $10$ GeV electron in a $10 \mu$G magnetic field is
$r_{g} \approx 3 \times 10^{11}$ cm.  Scattering lengths within several orders
of magnitude of this scale yield diffusion lengths smaller than the solar 
system and acceleration timescales $\lesssim 1$ yr at a typical fast shock.  
Dynamical scales, however, are generally measured in kpc and kyr, and often 
much larger scales.  

 	While this discrepancy makes solving equation \ref{convdiff} 
impractical with conventional computational methods, it can be used to 
develop a simplified and efficient electron transport equation, 
valid where $f(p)$ is 
sufficiently broad that it is adequately represented as a piecewise power 
law over finite momentum bins.  The key point here is that the comparatively 
miniscule length- and time-scales for electrons make this a natural 
assumption, since rapid diffusive shock acceleration for $\lesssim 10$ GeV 
electrons ensures that they will emerge from shocks with power-law momentum 
distributions effectively instantaneously compared to the overall dynamics.  
Subsequent cooling taking place downstream can be treated in a straightforward
way.  We therefore divide the momentum domain into a small number $N$ of 
logarithmically spaced bins, and estimate particle fluxes across momentum bin 
boundaries by representing $f(p) \propto p^{-q(p)}$ within bins, where $q(p)$ 
varies in a regular way.  Numerically, for each bin we use the total number 
of electrons within that bin and the associated mean logarithmic slope; the 
computational cost for a modest number of bins (we typically use $8$) is 
comparable to that required for tracking the dynamics, and therefore not 
prohibitive.  Standard finite-difference schemes for solving equation (1) 
would require more than an order-of-magnitude greater computational effort.
The scheme employed in the present simulations has been enhanced over the 
original version described in JRE99.  Most importantly, we have been able to
remove the constraint on the original scheme that limited the change in 
slope from one momentum bin to the next.  Thus, spectral cutoffs due to 
aging are handled better in the current scheme.  Otherwise, the methods are
the same as employed there.  Detailed tests were presented in that paper.

	In the test-particle limit for diffusive shock acceleration, electrons 
emerge from shocks with a power law spectral index $q=\frac{3r}{r-1}$ where 
$r$ is the shock compression ratio.  We allow injection of electrons from the 
thermal plasma at shocks using a model wherein a fixed fraction 
$\epsilon$ of the total electron flux through a shock is injected and 
accelerated to the appropriate power-law distribution beginning at momenta 
just above the postshock electron thermal values.  Here $\epsilon$ is set to 
$0$ for models 1 and 3, and $10^{-4}$ for model 2.  Electron injection at 
shocks is not well understood at present.  However, nonlinear models of strong
shocks usually lead to proton injection values $\sim 10^{-3}$ 
\citep[\eg][]{Gieseler00}.  Combined with the relative electron injection 
ratio of a few percent estimated by Bykov and 
Uvorov \citep{Bykov99}, a value of $10^{-4}$ in our simulation is very 
reasonable.  We note, of course, since the electron population in the 
simulations is passive, that results could be rescaled for an alternate
$\epsilon$.  In any case, if the jet flow consists mostly of ``thermalized''
plasma, our current understanding of collisionless shocks places 
$\epsilon > 0$. We include simulations both with $\epsilon = 0$ and 
$\epsilon >0$, since our purpose here is to explore the importance of
including this feature in the simulations.

	Our transport approach is complementary to that described by 
\citet{Micono99}.  There, the convection-diffusion equation was followed in 
detail outside of shocks using conventional numerical methods, but the 
calculation was constrained to follow a small number of selected Lagrangian 
volume elements in the flow because of high numerical costs. While that method
is not reliant upon the assumption of a piecewise power law momentum 
distribution, it does not make possible synthetic observations of the 
simulated flows, whereas ours does.  Another complementary approach is offered
by \citet*{Downes01}.  Those authors also treat the particle acceleration
as an injection process by specifying the fraction of the thermal
particle flux converted to nonthermal particles at shocks.  There the
Liouville equation is numerically integrated for particles accelerated 
in relativistic shocks, rather than solving the diffusion-convection 
equation directly.  
That method is computationally efficient; however it does not allow for 
self-consistent reacceleration of previously-accelerated particles 
(T. Downes 2000, private communication).  The use of a purely hydrodynamic 
code forces an assumption about the partitioning of post-shock thermal energy 
between energetic particles and the magnetic field in order to model
synchrotron emission.  Our work is restricted to the consideration of 
nonrelativistic shocks.  

\section{SIMULATED JET PROPERTIES} \label{jetprop.3.s}

  	Our simulated MHD jets described here are all dynamically identical.
Each enters the grid with a simple ``top hat'' velocity profile and a core 
speed that is Mach $80$ with respect to the uniform ambient medium ($M_{j} 
= u_{j}/c_{a} = 80$).  The high-velocity core is insulated from the ambient 
medium by a narrow, three-zone-thick transition layer.  The jets are
initially in approximate pressure balance with the ambient medium and have a 
density contrast $\eta = \rho_{j}/\rho_{a} = 10^{-2}$.  Thus the jet-based 
Mach number is $8$.  The jet enters at $x = 0$, with an initial core radius, 
$r_{j}$, of $15$ zones, while the entire grid is $576 \times 192 \times 192$
uniform zones ($38\frac{2}{5} r_{j} \times 12\frac{4}{5} r_{j} \times 
12\frac{4}{5} r_{j}$). Defining length and time in units of initial jet 
radius ($r_{j} = 1$) and ambient sound speed 
($c_{a} = (\gamma P_{a}/\rho_{a})^{1/2} = 1$, with $\gamma = 5/3$), the 
simulations are followed for $5.4$ time units, at which time the bow shock 
reaches the boundary at $x=38\frac{2}{5}$.  Because our objectives depend
on maintaining a fine resolution of the dynamical structures in the simulated
jets and their interiors, the large effort inherent in these simulations 
constrains us to look at relatively young flows for the time being.  
There is an initial axial background magnetic field ($B_{y} = B_{z}
= 0;~ B_{x} = B_{x0}$), with a magnetic pressure 1\% of the gas pressure 
($\beta = 10^{2}$). This results in an initial jet Alfv\'enic Mach number of 
$M_{Aj}=70$ compared to the hydrodynamic Mach 8 figure above.
In addition to the axial component $B_{x0}$, the inflowing jet also carries
a toroidal magnetic
field component representing a uniform axial current within the jet, with a 
return current on the jet surface; i.e., $B_\phi = 2 \times B_{x0}(r/r_{j})$ 
for $r \leqslant r_{j}$.  

	Open boundary conditions are used everywhere except at the orifice 
where jet material enters the grid.  Another commonly-used boundary condition
places a reflecting boundary outside the jet orifice at $x=0$ 
\citep[\eg][]{Norman96}. We adopt the open boundary at $x=0$ with an
eye towards the interpretation that the jet orifice is not a meaningful
model of the true jet origin, and instead represents where the collimated
jet enters our grid after having been launched further upstream ($x<0$).
\citet{Cox91} point out that the reflecting boundary at $x=0$ is not a
true representation of a source with point symmetry about the jet origin,
but is a very reasonable approximation for small precession angles.  

	As described below, we find very high
backflow velocities in the cocoon.  While
consistent with expectations based on general physical arguments, the backflow
structure might have been different had we instead used a reflecting boundary
at $x = 0$.  In that case weak shocks
might be more likely to form in the supersonic backflow.  The open boundary
near the jet orifice also contributes to the cocoon pressure and temperature 
gradients described in \S 4.1.  Because the oldest backflow material leaves 
the grid prior to the end of the simulation, the remaining cocoon gas has 
less time during which to come to equilibrium with the ambient medium.   
Nevertheless, our main conclusions are not significantly affected by the 
choice of boundary conditions.  

	\citet{Norman96} found in 3D jet simulations that even in the absence 
of an explicitly applied perturbation to the jet velocity, symmetry will 
eventually be broken by the action of Kelvin-Helmholtz instabilities on the 
jet, seeded by perturbations from pockets of supersonic turbulence in the
cocoon.  We found that a dynamically identical run to those presented
here differing only in a lack of precession remained essentially 
two-dimensional for sufficiently long times as to make this approach 
very expensive for study of three-dimensional behaviors.
Thus, to break cylindrical symmetry, we add a modest precession
to the in-flowing jet velocity.  

	In fact, evidence for precession is seen 
in a number of extragalactic sources \citep[\eg][]{Condon84,Kellermann88,
Mantovani99, Sudou00}.  Models for explaining the physical basis of 
precession in both galactic and extragalactic sources have been discussed in 
the literature \citep[\eg][]{Heuvel80,Spruit00}.  The jets presented here 
were precessed on a cone of opening angle $5\degr$, as in the first 
simulation presented in \citet{Cox91}.  The precession frequency was chosen 
so that the jet inflow velocity has just completed 5 revolutions by the end 
of the run.  
As discussed in section 4.1 below, we concluded that precession 
at this frequency was slow enough to avoid undesirable effects from 
``overspinning'' the jet such as excitation of undesirable stabilizing 
normal modes \citep{Hardee99}, while still breaking symmetry.  

	We present here three examples of electron transport within the
dynamics of the above jet flows.  Their properties are listed in Table 1.  
The essential character of these models was chosen to be analogous to the 
axisymmetric models explored in the earlier axisymmetric work of JRE99.  
Electrons are modeled explicitly in the momentum range $p_{0} < p < p_{N}$ 
with $p_{0} = 10~m_{e}c$ and $p_{N} \approx 1.63\times10^{5}~m_{e}c$ for all 
models.  Below $p_{0}$ the spectrum is assumed to continue as a 
simple power law, providing the necessary momentum-space boundary condition.  
The boundary condition at $p_{N}$ was set by assuming
$\frac{dq(p_{N})}{d\ln{p}}$ continuous.
Eight momentum bins ($N = 8$) were used for each of these models, 
yielding $\ln (p_{i+1}/p_{i}) = 1.5$ for all models.  Since the nonthermal 
cosmic-ray electrons are passive within our simulations, all results could be 
rescaled for different choices of $p_{0}$ ($p_{N}/p_{0}$ is fixed).  
All three models 
include the effects of adiabatic cooling and diffusive shock acceleration 
upon the cosmic-ray electron populations.  Second-order Fermi acceleration and
nonradiative energy losses such as Coulomb losses are neglected.

	In all three models, the nonthermal particle population entering the 
jet from the orifice comes with a momentum index $q = 4.4$, which represents 
a synchrotron spectral index of $\alpha = (q-3)/2 = 0.7$, where 
$S_{\nu} \propto \nu^{-\alpha}$.

	Models 1 and 2 are ``adiabatic'' models, in the sense that the 
electrons experience negligible synchrotron aging in these models, resulting 
in little spectral curvature.  Some curvature can result from spatial mixing, 
however. Models 1 and 3 are also similar to each other, since the electron 
populations in these models originate entirely within the in-flowing jet. 
The ratio of nonthermal to thermal particle number densities
in the jet, $\delta$, was set to $10^{-4}$ in these models.  
Model 2 is distinguished from the other models in that the electron population
is mostly injected locally from thermal plasma at shocks within the simulated 
flow ($\delta = 10^{-8}$, $\epsilon > 0$); this local injection of fresh 
particles was turned off 
($\epsilon = 0$) for models 1 and 3.  Model 3 differs from the others via 
significant radiative aging from synchrotron and inverse-Compton processes.

	In order to include the effects of synchrotron aging, the 
characteristic cooling time $\tau_{s0}$ must be defined in  
the computational time units, $r_{j}/c_{a}$.  Following the treatment
in JRE99, we express the cooling time in terms of the particle 
momentum and local magnetic field as
\begin{equation}
\tau_{s} = \tau_{s0}{\hat{p} \over p} = 25{1 \over p_{4}}{u_{j8} \over M_{j}}{1 \over r_{jk}}{1 \over (B_{10}^{2} + B_{cmb}^{2})}
\end{equation}
where $\hat{p}$ is a convenient fiducial momentum, $p_{4}$ is the electron momentum 
in units of $10^{4}~m_{e}c$, $u_{j8}$ is the jet speed in units of 
$10^{8}~cm~s^{-1}$, $M_{j}$ is the jet Mach number in the ambient medium,
$r_{jk}$ is the inflowing jet radius in kpc, $B_{10}$ is the field
strength in units of $10 \mu$G (nT), and $B_{cmb}$ 
($= 3.2\mu $G$~\times ~(1+z)^{4}$)
is the 
equivalent magnetic field of the cosmic microwave background at redshift $z$.
All of the models depicted here were chosen to have $M_{j} = 80$ as 
noted above, as well as $r_{jk} = 2$ and $u_{j8} = 15$, which corresponds to 
$0.05c$.  This makes the length of our computational box about $77$ kpc, and 
the time unit $r_{j}/c_{a} \sim 10^{7}$ yr.

 	Picking a fiducial momentum $\hat{p} = 10^{4}$ ($p_{4} = 1$), 
the cooling rate is then parameterized by the field $B_{10}$.  For models 1 
and 2, we have set $\tau_{s0} = 1.6 \times~10^{3}$ (compared to 5.4 as the
length of the simulation) by setting $B_{x0} = 0.39 \mu $G and ignoring the 
contribution from inverse Compton scattering off of the cosmic microwave 
background, the intent being to ensure negligible aging for electrons of 
interest within the length of the simulations.  

	On the other hand, for model 3 we set $B_{x0} = 5.7\mu $G, which 
corresponds to $\tau_{s0} = 5.4$.  The contribution from the cosmic 
microwave background is taken into account by setting $B_{cmb} = 3.2\mu $G 
corresponding to the current epoch.  Thus, for this model, aging effects are 
significant for the electron energy range of interest.  

	Note that the characteristic cooling time $\tau_{s0}$ depends very
strongly on the redshift, $z$, once Compton scattering from the cosmic
microwave background becomes important.  Thus inverse-Compton losses can lead 
to significant cooling in high-redshift sources, even if the magnetic fields
are relatively weak.  In fact for high $z$, it is 
difficult to scale simulations like these such that the nominal cooling time is
on the order of the simulation duration (as it is in model 3) without
making the jet speed $u_{j8}$ relativistic, as required by our code.

\section{DISCUSSION} \label{disc.4.s}
\subsection{Flow Dynamics} \label{flowdyn.4.1.s}

	Here we sketch out the key dynamical behaviors of the simulations, 
taking advantage of the fact that all three models are dynamically identical.
As expected for strongly-driven nonequilibrium systems \citep{Sato96} and
as has been seen in previous jet simulation work
\citep{Norman82,Williams85,Lind89,Hardee90,Cox91,Clarke96}, 
the dynamics is often characterized by very complex, nonsteady flows.  JRE99 
found that nearly all structures in their axisymmetric models were highly 
variable.  We find this to be equally true in the three-dimensional 
calculations as well, but while the end result (extreme variability of the
flow) is rather similar in 2D and 3D the underlying physical processes are 
quite different.  

	One of the most prominent characteristics of both two- and three-
dimensional jets is a seemingly chaotic, nonsteady jet termination shock.  In
both cases the jet terminus is so strongly perturbed at some times that it
is nearly impossible to identify a strong terminal shock at all.  In
two dimensions, this arises from the intermittent interaction between oblique
shocks within the jet, the Mach stem, and vortices shed from the jet.  This 
interaction gives rise to dramatic fluctuations in the size, strength, and 
location of the terminal shock, causing it to disappear almost entirely at 
some times, as well as inducing significant variations in the advance speed 
of the jet terminus.  The spectral distributions of the nonthermal particle 
populations processed by these shocks are correspondingly intricate.  In two
dimensions the role of the oblique shocks may be overemphasized by the 
axisymmetry, which acts to focus them upon the jet axis.

\begin{figure}
\begin{center}
\includegraphics{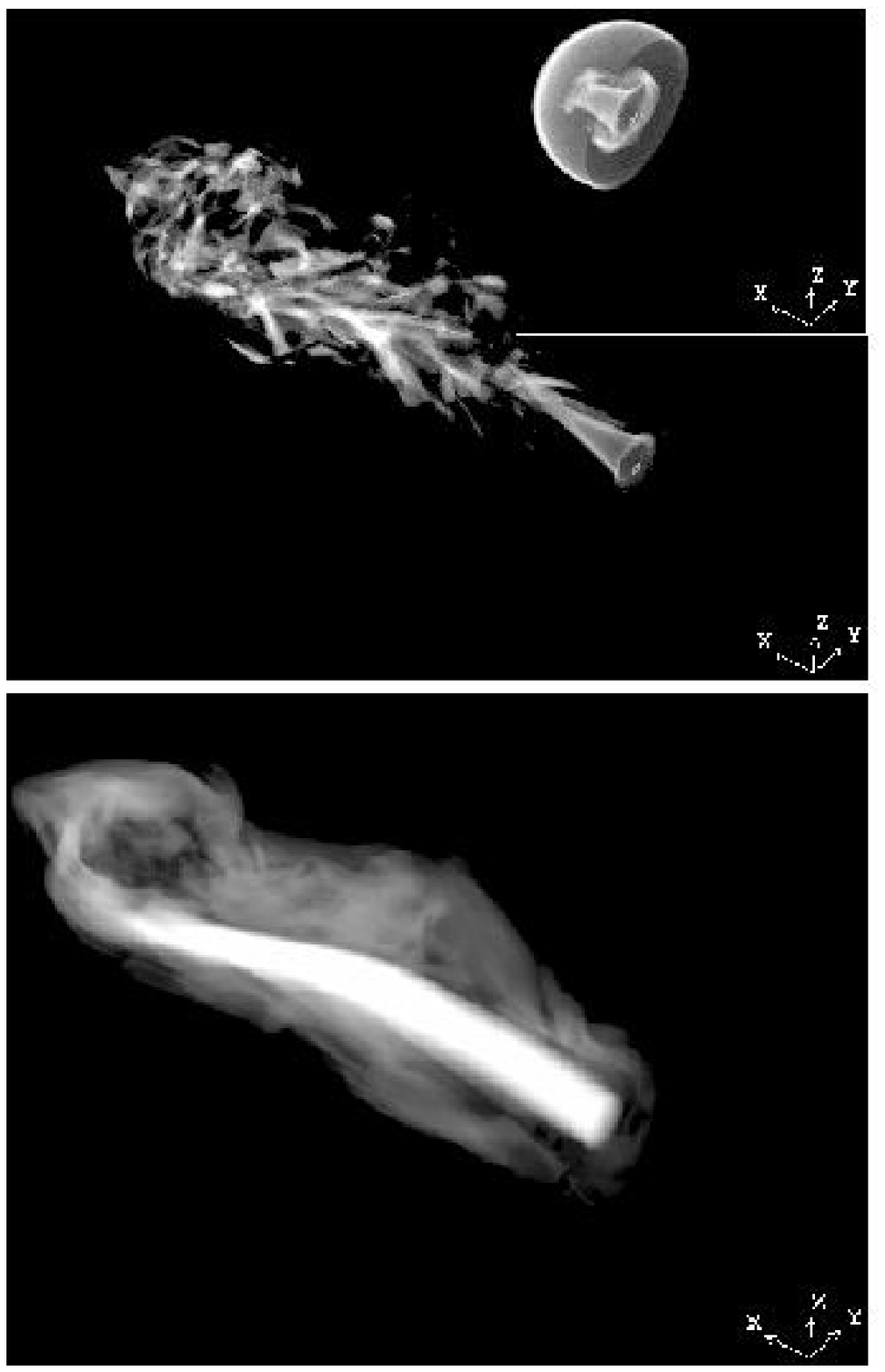}
\figcaption[f1-2.eps]
{(a) Top: Volume-rendered images of the shock structure 
($\nabla \cdot \mathbf u$)
in the jet flow.  The inset shows the situation at $t=0.7$, prior to breakup
of the terminal Mach disk, where the background has not been filtered out
in order to make the bow shock visible.  The main image shows the ``shock-web 
complex'' at $t=4.0$, for zones containing only jet-supplied material 
($C_{j} \geqslant 0.99$). The small square in 
the lower right marks the jet orifice, and the orientation corresponds to the 
jet pointing slightly out of the page.
(b) Bottom: Volume-rendered jet velocity magnitude ($|\mathbf u|$) 
at $t=4.0$.  
The orientation is the same as in (a), and ambient material has been filtered
out ($C_{j} \geqslant 0.99$).
\label{shocks-vel.f}}
\end{center}
\end{figure}

	 In three dimensions the nonsteady nature of the terminal shock is 
even more profound.  At the earliest times, the jet structure is nearly 
axisymmetric, with a clearly defined Mach disk terminating the jet flow and
a nice symmetrical backflow (figure \ref{shocks-vel.f}a, inset) 
as has been seen in even the 
earliest two-dimensional calculations \citep[\eg][]{Norman82}.  
The Mach disk is disrupted rather abruptly once the first set of oblique 
shocks converge upon the jet axis \citep{Clarke96} which
occurs at approximately $t=1$ in our time units, corresponding both to one
sound-crossing time and coincidentally to one complete revolution of the 
precessed inflow velocity.  Unlike the two-dimensional case, however, the 
Mach disk is never recovered after its initial disruption.  This is very
easily seen in the relevant animation found on our web site
\footnote{\url{http://www.msi.umn.edu/Projects/twj/radjet/radjet.html}}.  
Present on this site are volume-rendered animations of velocity divergence, 
velocity magnitude, and magnetic pressure.  Many of the dynamical behaviors 
detailed in this section can be viewed directly in these animations. 

	From that point on, the terminal structure of the jet is better 
described as a ``shock-web complex'' rather than as a simple strong terminal
shock (figure \ref{shocks-vel.f}a).  This complex encompasses a 
network of shocks 
of varying strengths and sizes spread throughout the source, including the 
head region and backflow.  Rapid spatial and temporal changes in the structure
of the complex attest to the fleeting nature of many shocks in the system.
At the times when the end of the jet truly is associated with a simple 
terminal shock, that shock is typically too small to capture most of the 
material passing from the jet into the cocoon.  Furthermore, the strongest 
shock in the system is often not the terminal shock.  Therefore, in these
simulated flows it does not make sense to describe particle acceleration
predominantly in terms of the canonical jet terminal shock.  
Our code accurately captures strong shocks within 2-3 computational zones
\citep{Ryu95}, so we are confident that the variety of shock strengths is not
an effect of numerical viscosity.  These methods do not employ a numerical
viscosity, in fact, so shock jumps are quite accurate.

	While still present in three dimensions, the oblique shocks
within the jet exercise far less control over the overall jet dynamics. 
In addition, we find several other kinds of shocks arising at different times 
and locations within the flow. Precession (or more generally any terminal
jet ``wobble'' such as that which arises naturally from fluid instabilities) 
plays an important role here.  At some times the terminal jet velocity is 
oriented such that the beam ``splashes'' or 
``splatters'' against the contact discontinuity that forms the cocoon 
boundary. In these instances the jet impinges on the cocoon boundary at an 
angle less than the maximum bending angle (the angle beyond which bending
causes disruption of the jet \citep[eq. 5.58]{Icke91}) and is redirected 
supersonically back into the cocoon, occasionally even impinging upon the 
far wall, as has been seen previously by  \citet*{Williams85} and 
\citet{Cox91}. These ``jet redirection events'' are visible in images of 
velocity divergence as intermediate-strength shocks on the cocoon boundary 
and as magnetic field enhancements in images of magnetic pressure. At other 
times, the combined actions of wobble and fluid instabilities cause 
the jet (as identified by a relatively 
well-collimated high-velocity core; see figure \ref{shocks-vel.f}b) to develop 
severe bends, sometimes in 
excess of the maximum bending angle.  Examination of purely hydrodynamical 
considerations places this value at roughly $70 \degr$. In these cases the 
bending is a precursor to ``breaking'' and realignment of the jet flow.  
When this happens the portion of the beam downstream of the ``elbow'' is 
disconnected from the main flow and deposited in the backflow; lacking an 
energy source, this flow quickly dissipates. (This, in fact, is one mechanism 
for introducing relatively large amounts of new jet material quickly into the 
backflow.  It also provides an energy supply for some of the radio hotspots 
that occur in the synthetic observations, as discussed in section 4.3.1.)  
Following disconnection, this ``elbow'' itself produces
a strong, transient terminal shock:  the new end of the jet is 
now straight and pointed roughly towards the end of the cocoon rather than 
the edges. Note that the terminal shock forms in this case after the bending 
has become significant, and thus the material downstream of the elbow can 
enter the backflow without being processed via the shock.  The temporary 
reappearance of the terminal shock is accompanied by the rapid extension of 
the lobe as seen in the ``dentist's drill'' model of \citet{Scheuer82}, and 
the creation of a narrow ``finger'' that pokes out of the contact discontinuity
where the jet surges forward, resulting in a corresponding deformation and 
temporary strengthening of the bow shock.  In fact, the bow shock is 
occasionally among the strongest shocks in the system during these events.  

	In contrast to the situation in three dimensions, instances of 
disruption and reformation of the terminal 
shock in two-dimensional simulations were coincident with distinct episodes
of strong vortex shedding.  In these episodes large ``rolls'' of material
are thrown into the backflow, where they interact with the Kelvin-Helmholtz
unstable boundary layer of the jet, further perturbing the jet flow.  
Axisymmetry in this case forces the shed vortices to be strictly annular,
thereby artificially enhancing their effect on the jet.  In three dimensions,
the vortex structures are stretched and tangled more completely, leading
to turbulence and disorder on smaller scales, in turn giving rise to vortex 
tubes.  The turbulent cascade of energy and disorder to smaller scales in the 
three-dimensional calculation causes instances of extended structures
in the backflow acting coherently on the jet to be more unlikely than
in two dimensions.  On the other hand, the enforced axisymmetry of the 
two-dimensional calculation allows the jet to experience only pinching modes, 
such as those driven by shedding of the annular vortices, while the 
three-dimensional calculation makes the jet susceptible to a much 
larger range of possible dynamical modes, including pinching, bending, and 
twisting modes \citep*[\eg][]{Hardee97}.  While not as effective as the 
asymmetric backflow in slab-symmetric calculations \citep*[\eg][]{Hardee90},
the turbulent three-dimensional backflow does contribute strongly to the
``flapping'' of the jet by providing perturbations that seed these 
unstable modes. Instances where large amounts of jet material are quickly 
introduced into the backflow as in the vortex shedding events in two 
dimensions are not seen as frequently in three-dimensional calculations, 
although the previously mentioned ``breaking'' of the jet terminus can have 
a similar effect.

	As the beam grows in length, precession at
the inflow boundary is transmitted down the jet as increasingly larger
changes in the velocity direction over the precession period, further 
contributing to the ``flapping'' of the jet, particularly at the latest times.
As mentioned above, earlier simulations \citep{Norman96} have shown that 
three-dimensional hydrodynamical jets ($M_{j} = 100, \eta=10^{-2}$) will 
eventually begin to flap in the absence of precession.  
There the timescale for side-to-side deflection of the jet occurred
on the internal dynamical timescale, $r_{cocoon}/c_{cocoon}$, which in that 
case was $4-5$ times shorter than the time for the jet head to advance one 
cocoon radius, $r_{cocoon}/v_{head}$.  The precession-driven flapping which 
takes place in the dynamical jet model described here is a bit slower than 
this.  Consideration of the contact discontinuity near the head region 
places the effective cocoon width at the end of the 
simulation as half the grid width (although in reality the size of the head 
region varies considerably).  Thus by this time the jet head has advanced 
roughly $5.5$ cocoon radii.  As mentioned above, the period of the applied 
precession is such that the jet has just finished $5$ revolutions by the end 
of the run, or approximately $10$ side-to-side motions.  The result is that 
the timescale for side-to-side motions of our simulated jet is only about a 
factor of $2$ shorter than that for the jet head to advance one cocoon radius,
compared to the factor or $4-5$ that arises from instability-induced 
perturbations in Norman's model.  The precession applied here is therefore
``gentler'' than the flapping that can arise naturally from hydrodynamic 
instabilities in simulations of jets similar to that modeled here. 

	Although jet breaking events are accompanied by extension and
strengthening of a portion of the bow shock, the overall advance of the 
jet head is determined by the momentum flux transported down the jet,
averaged over the effective working surface of the entire head region.
The average advance speed of the bow shock across the grid is consistent
with that predicted by \citet*[][(equation 1)]{Frank98}, where 
$\alpha \approx \onethird - \onequarter$ is used to approximate the 
averaging of the jet momentum flux over the head region to account for
lateral motions of the jet.  (Without averaging, $\alpha \approx \frac{1}{6}$.)
When discussing the advance of the radio lobe, one must decide what
feature, either in the dynamics or the emissions, will be used to 
define the lobe.  Here we have chosen to consider the advance speed of
the jet head in terms of the bow shock.
Since the ambient medium is uniform here, we expect the lobe advance speed to 
decline as the effective area of the head region increases 
\citep{Cox91,Falle91}.
The size of the computational grid used here caused the simulations to end
before such longer-term jet behaviors could manifest themselves.
Precession and jet breaking events cause the advance speed of the radio 
hotspots, when they are visible in the synthetic observations, to vary 
significantly.

	We find that backflow velocities in excess of $30\%$ of the initial
jet velocity ($v_{b} \gtrsim 25$ in simulation units) are common, sometimes 
reaching up to  $50\%$ of the jet speed in isolated instances 
(figure \ref{shocks-vel.f}b).  A reflection
symmetry at $x = 0$ might have reduced the backflow rate, however.  The 
backflow is host to turbulence that is driven not only by fluid instabilities 
on the jet/backflow and cocoon/medium boundaries, but also by lateral motions 
of the precessing jet and especially its ``flapping''.  The flows are highly 
nonsteady.  Nonetheless, velocities in the regions of strong backflow are 
qualitatively consistent with those estimated by the Bernoulli equation for 
flow between the jet head region and the cocoon in an axisymmetric source 
\citep{Norman82}. Since the Bernoulli equation is a statement about 
steady-state flows, the application to this source is not rigorous.  The 
qualitative agreement that we find between the backflow velocities in our 
model and those predicted via the Bernoulli equation may be indicative of the 
long-term time-averaged behavior of the cocoon.
Away from the head region, the flow is typically only mildly supersonic or 
trans-sonic, with internal Mach numbers often less than unity.  Transient 
strong shocks are occasionally observed in the backflow, to an extent 
that they can have a discernible effect on the nonthermal electron population 
there.  This is particularly true closer to the head region where 
internal Mach numbers in the backflow are slightly higher, as would be
predicted from a model where supersonic turbulence close to the working
surface decays as the backflow passes out of the head region \citep{Norman96}.
  
	Similarly fast backflows are seen in the two-dimensional simulations 
of JRE99.  There the backflow velocities ranged over $30\% - 50\%$ of the 
initial jet speed along the length of the computational grid.  In that case 
the backflow was compressed and reaccelerated through a De Laval-type nozzle,
as the imposed symmetry allowed the subsonic plasma in the head region only a 
small channel to escape from the head into the backflow around the Mach stem.  
Although such channels may be temporarily disrupted during episodes of strong
vortex shedding, 
we expect this mechanism to be much more prominent in the two-dimensional case
since the Mach disk exists only at earliest times in the three-dimensional 
simulations presented here.  While expansion and reacceleration of the flow 
in the cocoon may still be the origin of the supersonic backflow found in the
three-dimensional case, such channels are constantly created and destroyed
by the effects of precession and jet flapping.

\begin{figure}
\begin{center}
\includegraphics{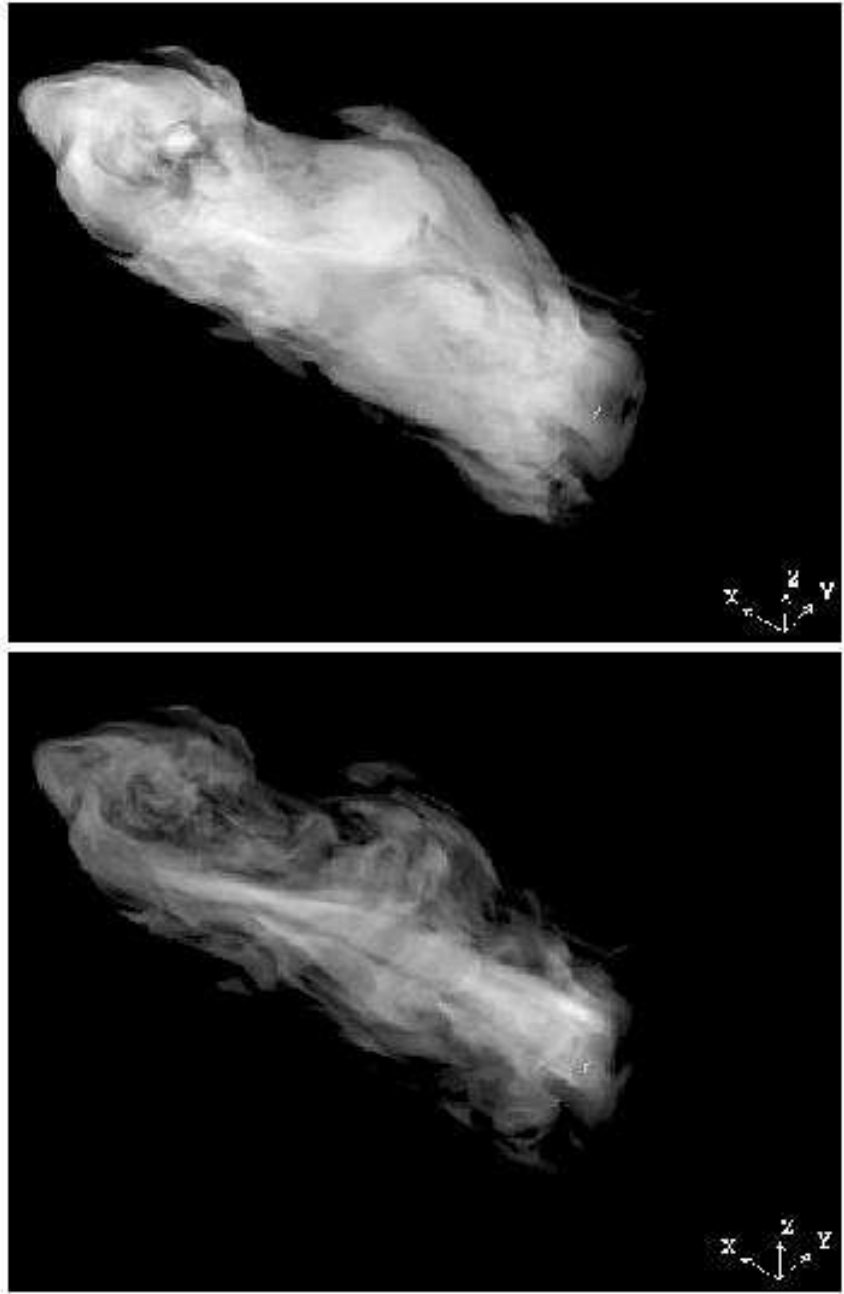}
\figcaption[f2-2.eps]
{(a) Top: Volume-rendered image of log of thermal plasma pressure at $t=4.0$.  
Displayed pressures span a range of $10^{2}$.
(b) Bottom: Volume-rendered image of log of magnetic pressure at $t=4.0$.  
Displayed pressures span range of $10^{3}$.  Most of the largest values are 
found in the jet core. Large volumes of the cocoon are occupied by magnetic 
pressures more than two orders of magnitude below the nominal value in the jet.
In both images, only material with $C_{j} \geqslant 0.99$ is shown.  
\label{pressures.f}}
\end{center}
\end{figure}

	The considerable decrease of the internal Mach number of the flow
between jet and cocoon demonstrates that the plasma is strongly heated by
compression in the head region, despite the absence of a persistent terminal 
shock.  Thermal gas pressure in the head region is comparable to the momentum
flux density transported down the jet, independent of whether or not a 
terminal shock exists.  As expected, the cocoon is overpressured with 
respect to the jet
\citep[\eg][]{Begelman89}.  While the gas pressure is 
generally larger near the jet head, the pressure distribution in the cocoon
is quite complex (figure \ref{pressures.f}a). This reflects the violent
dynamics described already.   
Pressure values around the head region generally approach and in some places 
exceed a factor of $5$ over the original pressure in the jet and ambient 
medium.  This ratio can be much higher (up to a factor of nearly $50$ in one
case) in small regions where the jet impinges directly on the contact 
discontinuity, such as after a ``breaking'' event.  This is consistent 
with the expectation that the pressure just downstream of a strong terminal
shock should increase by roughly the square of the internal jet Mach number, 
or about $64$ in this case.
The overpressure is not 
nearly as dramatic further back in the cocoon, and rarely exceeds a factor 
of $2$ near the jet orifice.  
 
	While the jet is underpressured with respect to the cocoon, not 
surprisingly it is overdense with respect to the cocoon.  The cocoon density
does not have the same clear gradient as the pressure.  The density 
generally varies between one-half and one-third of the jet density throughout 
most of the cocoon, although specific values fluctuate quite a bit from
location to location.  Some of these fluctuations bring the cocoon density
close to the density of the jet downstream of the first oblique shock, or
about $1.5$ times the initial density (giving the first oblique shock a
Mach number of $1.3$).  In a thin layer just inside the contact discontinuity,
the cocoon
density is typically closer to this jet value downstream of the oblique 
shock rather than the much lower densities found throughout the rest of the
cocoon.  
The low density and high pressure of the cocoon contribute 
significantly to the overall stability of the jet.  As is generally the
case for heavy jets, the overdense jet propagates ballistically through
the cocoon.  

	The temperature distribution in the cocoon is naturally also 
very complicated.   A temperature gradient is discernible in the backflow,
however, as adiabatic expansion works to cool the plasma while it flows 
towards the core. As expected in light of the fast backflow velocities just 
described, the backflow near the jet orifice has not achieved 
equilibrium with the ambient gas by the end of our simulation.  The age of
oldest material in the backflow near the orifice is only roughly half the 
simulation duration by the end of the simulation, as older material 
has already 
exited the grid via the open boundary.  The plasma beta parameter increases 
significantly ($\beta \sim 10^{3}-10^{4}$) throughout much of the cocoon, 
owing to the combined effects of the increased thermal pressure and the 
decreased magnetic fields described below.  Overall we see little detailed 
similarity
between the distributions of thermal pressure and magnetic pressure in the
cocoon plasma, as shown in the volume-rendered images of thermal and magnetic
pressure in figure \ref{pressures.f}.  This is expected, since the former is 
generally enhanced by compression, whereas shearing flows are often the cause 
of magnetic pressure increases, as described below.  

	Alfv\'enic Mach numbers in the jet are initially close to $70$.  In
conjunction with the plasma beta parameter $\beta \sim 10^{2}$, this indicates
that the jet magnetic field is dynamically unimportant at the outset of the 
calculation.  Alfv\'enic Mach numbers within the jet itself remain steady 
virtually all the way to the head region, with $\lesssim 25\%$ variations 
where weak oblique shocks temporarily compress the magnetic field.  Thus 
magnetic tension has a minimal effect upon the dynamics of the jet itself 
throughout the calculation.

	Magnetic field structures in the backflow are subjected to the same
stretching and twisting effects in three dimensions as the vortex structures,
so we expect to see the development of magnetic flux tube complexes as well.
The resulting nonuniformity can be expressed through the magnetic field 
intermittency, $I_{B} = \frac{<B^{4}>}{<B^{2}>^{2}}$ \citep*{Ryu01}, which can 
be crudely thought of as a reciprocal ``filling factor''.  Inside the jet 
lobes, $I_{B}$ can approach large values $\lesssim 100$ at late times, which 
indicates that the field has developed a predominantly ``bundled''  topology. 
This leads to the appearance of bright emission filaments in the cocoons of 
these objects when synthetic radio observations are conducted, as discussed in
section 4.3 below.  Such filaments are commonly observed in real radio 
observations 
(\eg M87 (\citealt{Hines89}), Cygnus A (\citealt{Carilli91}))
and are typically interpreted as magnetic structures.  We would agree, 
based on our simulations.

	Field-line stretching within shearing flows can quickly amplify 
magnetic fields, although such amplification does not appear to be sustained 
locally over long distances or times in the backflow.  Instead, regions of 
the most 
extreme field amplification are discrete and appear briefly.  (That
is not to say that their effect on the spectral distribution of nonthermal 
particles is quite limited. We will see in section \ref{radage.4.3.2.s} 
that they can 
give rise to large islands of very strongly aged material.)  Jet material
downstream of the ``elbow'' during a jet breaking event can get sheared by the
backflow, thus introducing small high-field filaments into the backflow.
Such instances can bring the field strength up to dynamical importance, 
although these filaments quickly undergo adiabatic expansion.  
Once the flow becomes truly three-dimensional, the mean magnetic
field varies slowly with an overall decrease by a factor of a few
over the duration of the simulation.  
This indicates that
large-scale field amplification such as that from a precession-driven
turbulent dynamo is not taking place.  
We would not expect a dynamo-driven amplification of the large-scale
fields, despite the apparent introduction of a net helicity via precession.
In fact it is important to point out that precession adds little or 
no net helicity at all to the problem, since in the absence of interactions
with the backflow, all jet material moves in a straight line from the
orifice.
Compression of 
the field where the beam splashes against the contact discontinuity is another
cause of transient field growth.

	The absence of axisymmetry-enhanced vortical rings in the 
three-dimensional calculations results in a greatly lessened tendency 
towards ``flux expulsion'' \citep{Weiss66}; that is,
the annihilation of poloidal field components driven by reconnection.  
Nevertheless large volumes of the cocoon are host to magnetic pressure values 
greatly diminished below the nominal value transported down the jet, often by 
two orders of magnitude or more.  This is consistent with the observation by
\citet{Clarke96} that weak lobe magnetic fields are typified by extended 
filamentary structures and large magnetic voids.  This is also consistent
with the large values of $I_{B}$ mentioned above.  Adiabatic expansion is the 
primary engine for this field reduction.  These large regions may act as 
``freezers'', tending to preserve particle spectra against continuous 
radiative aging \citep{Scheuer89,Eilek97}, if the energetic particles remain
in these volumes for long periods of time \citep{Blundell00a}.

	The diminishment of magnetic fields in the cocoon also drives the 
Alfv\'enic Mach number, $M_{A} = u/v_{A}$, up by an order of magnitude over 
its original value in the jet, $M_{Aj} \approx 70$, signifying that the 
plasma kinetic energy 
density exceeds the magnetic energy density and that magnetic tension is 
dynamically unimportant in these volumes.  However the magnetic filaments 
described above occasionally correspond to $M_{A} \lesssim 10$, 
which is nearly an order
of magnitude less than the original jet value.  The dissolution of these 
filaments is partially mitigated by the resident fields, although the 
Alfv\'enic Mach number rarely becomes much smaller than $10$. In contrast, 
JRE99 found many locations in their analogous axisymmetric models where 
$M_{A} \lesssim 1$.  There the flows were made smoother by the presence of 
the field, via reorganization and realignment of the flow after reconnection 
events. 

	One must use caution while interpreting detailed magnetic field
properties in these or any such simulations.  First, there are fundamental
differences expected between axisymmetric field evolution and fully 
three-dimensional fields.  For an axisymmetric field the $B_{\phi}$ 
component is partially decoupled from $B_{r}$ and $B_{z}$, so is not subject
to reconnection.  On the other hand $B_{\phi}$ can be strongly enhanced
by flux stretching if a plasma element expands in the radial direction.  
Thus $B_{\phi}$ tends to become dominant in axisymmetric MHD simulations.
That effect is largely gone in three-dimensional flows that are disordered,
since all components participate in the reconnection.  On the other hand flux
tubes in 3D tend to be stretched and twisted, so can become locally 
strengthened beyond two-dimensional flows \citep[see, \eg][]{Ryu01}.  At
the same time these effects are limited by finite numerical resolution, so
simulations are not able to establish accurately the smallest values of $B$
or $M_{A}$ that would be expected to occur locally.  We are confident, however,
that the global properties of the cocoon fields seen in these simulations
are correct.

	A similar comment holds regarding magnetic reconnection in these 
simulations.
The dissipative effects which give rise to reconnection here are purely
numerical in origin, and thus not representative of the microphysics that
governs such transitions in the real world.  These simulations are, therefore,
unable to provide information about the local rate of reconnection or the 
role of reconnection in heating the plasma in radio galaxy lobes.  Since
reconnection is basically a topological transition, however, its presence
in the complicated magnetic field configurations found in
our simulated sources is reasonable \citep[\cf][]{Ryu01}.

\subsection{Electron Transport} \label{transp.4.2.s}
	
	We now turn to a consideration of the electron transport effects 
within each model.  The cosmic-ray electrons in these simulations are 
dynamically passive, and in fact as a consistency check we note that 
their nominal energy density never exceeds
more than $0.1\%$ of the kinetic energy density anywhere in the computational 
grid.  Because the electron transport takes place ``on top'' of the dynamics, 
the fact that the three models outlined here are dynamically identical enables
us to examine in a straightforward way the means in which various 
momentum-transport effects can alter electron populations subjected to the
same dynamical history.  

\subsubsection{``Adiabatic'' Models 1 and 2} \label{transp-adiab.s}

	The absence of radiative cooling and injection effects in model 1
make it the most straightforward to analyze.    
Nevertheless the nonthermal particle populations in this simplest model 
deserve close scrutiny.  

	As stated previously, nonthermal electrons in this model enter the
grid via the jet orifice.  The negligible radiative cooling rate in this
model ensures that the particles are transported down the jet with momentum 
indices virtually unchanged, and the distribution of magnetic fields is
largely immaterial to this discussion.  Thus they enter the shock-web complex
as a uniform power-law momentum distribution with index $q=4.4$ before 
passing from the jet itself into the cocoon.  The oblique shocks along the
jet are too weak to modify the electron population in this model.

	Recognizing the role of the shock-web complex is crucial to 
understanding how the dynamics of these simulated sources affects the 
nonthermal electron transport and in particular the distribution of momentum
indices in the cocoon.  All material that enters the cocoon does so only
after passing through the diaphanous shock web, which may mean passing 
through a strong shock, passing 
through a weak shock, or flowing through a ``hole'' where no shock is found.  
In any case the shock web makes an indelible impression upon the manner in 
which the particle transport properties manifest themselves in the cocoon, 
because even in the control model (model 1) it allows a subtle variety of 
particle spectra to enter the cocoon.  

	In model 1, the only momentum transport effect to
which the particles in the shock web may be subjected is a flattening at
sufficiently strong shocks.  Thus only downstream of relatively strong (and
therefore somewhat rare) shocks with compression ratio $r \gtrsim 3.14$ 
will the particle  momentum index
differ from $4.4$.  Another way to express this is in terms of the shock
Mach number, which must exceed $M = \sqrt{q/(q-4)}$ to reaccelerate particles 
with momentum index $q$, and thus requires $M \gtrsim 3.3$ to flatten a 
distribution with $q=4.4$.  Recall that in many places in the cocoon, the 
internal Mach number is often less than unity.  Since there often is no 
terminal shock, and since when it does exist it is often weaker than 
$M \approx 3.3$, most of the jet material enters
the cocoon without being flattened.  Small amounts of flatter material are
sporadically introduced into the cocoon, at times and places  where a portion
of the flow passes through a sufficiently strong shock.  Considerable
variety is seen even in these strong shocks, so these small amounts of
shock-accelerated material enter the cocoon with an unpredictable variety of
momentum indices less than $4.4$.  Thus we see mostly momentum indices of 4.4 
in the cocoon, with small ribbons of flatter material downstream of strong
shocks.  The fact that we see spectral structure, even in the control model, 
serves as a reminder that by looking at momentum indices
we are not seeing a snapshot of the dynamical state of the system so much 
as we are seeing information about the dynamical history of the nonthermal 
particles.  

\begin{figure}
\begin{center}
\includegraphics{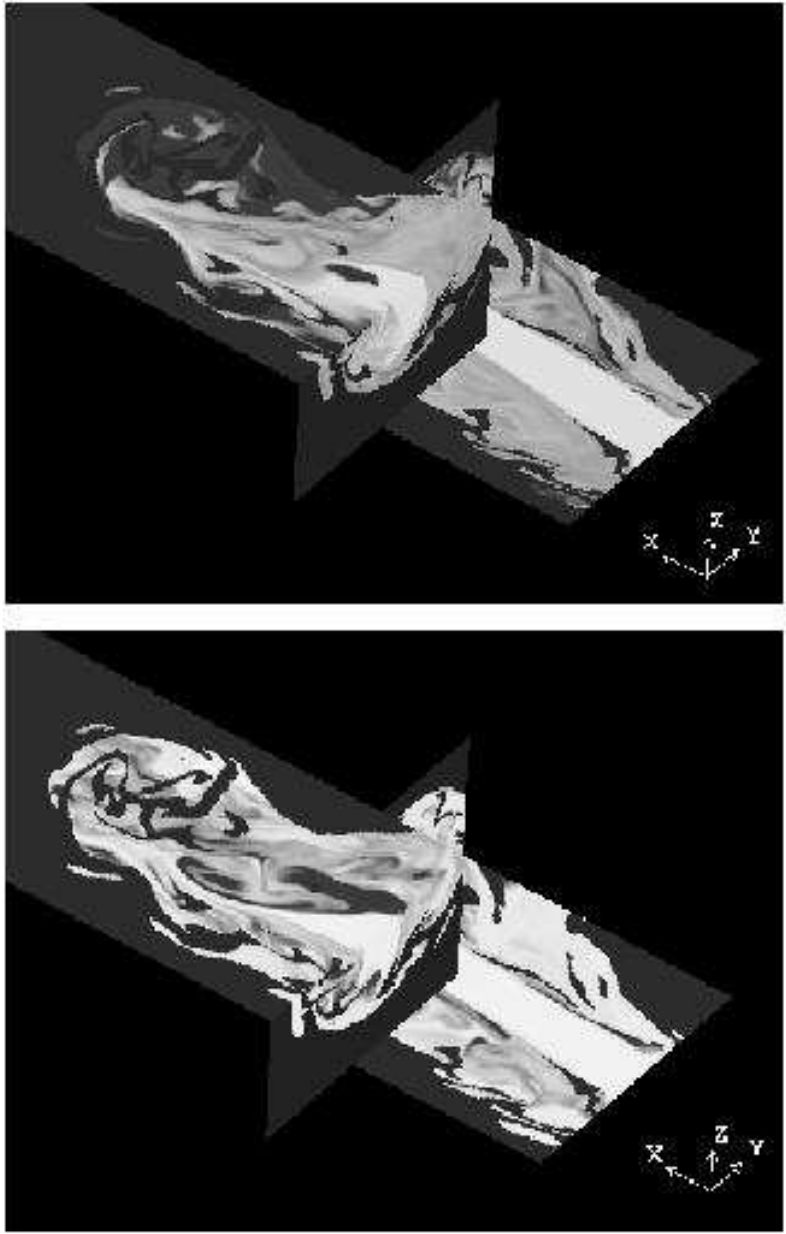}
\figcaption[f3-2.eps]
{(a) Top: Orthogonal slices through the computational grid, showing nonthermal 
electron momentum indices $q$ for electrons with $p \approx 3 \times 10^{4}$ 
in model 2 at $t=4.0$.  Notice the variegated spectral structure of the cocoon.
As usual, electrons entering with the jet have $q = 4.4$, but now 
fresh-particle injection at the myriad weak shocks in the shock web leads to 
an excess of steep-spectrum electrons in the head region. The display range is
$4 < q < 7$, with the flattest material light.
(b) Bottom: Corresponding image for model 3. The spectral structure of the 
cocoon is quite patchy, lacking both the uniformity of model 1 and the overall
gradient of model 2.  Here the display range is larger, with $4 < q < 9$ to 
capture the effects of radiative aging.  In both images only material with 
$C_{j} \geqslant 0.99$ is shown.
\label{qdist.f}}
\end{center}
\end{figure}

	In the injection model (model 2) as in model 1, it is predominantly 
through the action of the
shock web that the simulated dynamics affect the relativistic particle
transport.  Again the variation of magnetic field strengths throughout the
simulated source is essentially unimportant to the electron population because
of the extremely low radiative cooling rate in this model.
Yet the inclusion of fresh particle injection at shocks makes a profound
difference.  We found in the control model that basically only the strongest
shocks in the shock web influenced the particle transport and that the spatial
and temporal intermittency of such shocks limited their influence greatly.  
As before, in model 2 the
momentum indices flatter than the original jet value of $q=4.4$ are only
found downstream of the strongest shocks.  Yet unlike model 1, most of the
nonthermal electrons in this simulation are introduced at shocks rather than
at the jet orifice.  Thus, now all of the shocks in the shock web, weak and
strong alike, contribute.  In fact it is the weak shocks ($r < 3.14$, $q>4.4$)
that take center stage in this model.  Since internal Mach numbers in the
backflow are often only mildly supersonic as described above, weak shocks
outnumber the relatively strong shocks, ensuring that more steep-spectrum 
material enters the cocoon than flat-spectrum material.  This is particularly 
true at times corresponding with a ``jet breaking'' event, when the shock web 
is especially complex and spread over the entire head region.  This is 
demonstrated in figure \ref{qdist.f}a.  There we see predominantly more
steep material in the jet head (reflecting local injection by weak shocks)
than further back in the cocoon, near the jet orifice (reflecting the stronger
shocks formed early when the jet flow was almost two-dimensional).

\subsubsection{``Cooling'' Model 3} \label{transp-cool.s}

	When radiative aging becomes significant, the distribution of
magnetic field strengths throughout the simulated flow is no longer immaterial
to a discussion of the electron transport.  Thus in addition to 
the action of the shock web, the magnetic fields provide a second crucial
link between the source dynamics and the particle transport in model 3.

	The shock web plays more or less the same role here as it did in the
control model (model 1), with a slight difference.  Once again all of the
nonthermal particles enter the grid via the jet, but now radiative aging is
significant enough in this model that the relativistic electrons propagating
down the jet exhibit
a modest amount of spectral curvature by the time they encounter the shock
web.  Sufficiently strong shocks will reaccelerate the curved distributions
into power laws before they pass into the cocoon.  Yet as before a large 
fraction of the 
material coming down the jet passes through weak shocks, or indeed no shock
at all, before entering the cocoon.  Not only does the jet itself 
steepen noticeably between orifice and the terminal shock, which does exist
at this time, but the lack of 
fresh-particle injection in this model means that there is a paucity of 
flat-spectrum electrons at the post-shock location.   In the injection model, 
there was a significantly flatter population downstream of the terminal shock,
but not  here.  Most particles enter the cocoon without the benefit of strong 
shock acceleration.  Thus in contrast to models 1 and 2, we 
now see the introduction of non-power-law distributions into the cocoon even 
before radiative cooling can take place in that complicated magnetic 
environment.  

	As described above in section 4.1, large volumes of the cocoon are 
occupied by ``freezers'', where the magnetic pressure drops significantly 
below the inflowing jet values.  The radiative lifetime of 
nonthermal particles at the fiducial momentum inside these volumes exceeds the
nominal lifetime by more than two orders of magnitude.  Regions of rapid 
cooling are restricted to a smaller fraction of the cocoon volume, in the form
of the high magnetic field filaments that are generated by shearing flows at 
the jet head, and that thread the larger magnetic voids.  The extreme 
variability of the cocoon magnetic field, both spatially and temporally, makes
the creation of these strong-cooling regions sporadic.  Often we see different
regions within the same source between which the effective cooling rate can 
vary by orders of magnitude.  
We note however, that this would not necessarily be the case if the 
calculation were performed at a higher redshift.  As noted above,
the $(1+z)^{-4}$ dependence of the nominal cooling rate $\tau_{s0}$ means
that inverse-Compton losses become significant at high redshift even when
the relevant magnetic fields are weak. 
In model 3, the cooling contribution from inverse-Compton losses would
equal the contribution from synchrotron losses in the fiducial field 
($5.7 \mu$G)at a redshift of only $z = 0.16$.  Inverse-Compton losses 
would therefore dominate cooling in the magnetic voids at even lower
redshifts.

	The effects of radiative aging on the 
distribution of momentum indices in the source is seen in figure 
\ref{qdist.f}b. Also clearly visible in this figure are dark channels 
representing regions where ambient material has been entrained into the jet 
cocoon.  This illustrates the Kelvin-Helmholtz unstable character of the 
cocoon boundary, which leads at later times to mixing between the arbitrarily 
steeper ambient population and the flatter jet populations.

\subsection{Synthetic Observations} \label{synthobs.4.3.s}

	Our goal here is to develop an understanding of the connections
between physical structures and observable emission patterns present in these 
simulations, with eventual application to real radio galaxies. We emphasize,
however, that the current simulations are still intentionally idealized to
allow the best isolation of clear cause and effect relationships.
We utilized the combined vector magnetic field structure and nonthermal 
particle distributions within the simulations to compute a large set of 
synthetic observations of our simulated radio galaxies.  By synthetically 
observing a source whose detailed physical structure is known beforehand, we 
hope to gain insights into what real observations are reliably telling us.  
The simulated objects represent truly three-dimensional objects with a 
self-consistent particle energy distribution, so these synthetic observations 
are a big step beyond previous calculations of this type.

	In every zone of the computational grid we compute a
synchrotron emissivity \citep*[JRE99;][]{Jones74}
\begin{equation}
j_{\nu} = j_{\alpha 0}{4\pi e^{2} \over c}f(p)p^{q}\left({\nu_{B_{\bot}} \over
 \nu}\right)^{\alpha}\nu_{B_{\bot}}
\end{equation}
where $\alpha = (q-3)/2$; $j_{\alpha 0}$, a function of $\alpha$, is of order 
$1$ \citep{Jones74};  and $\nu_{B_{\bot}} = \nu_{B}~\cos~\theta$, the electron
cyclotron frequency in terms of the magnetic field projected on the sky plane,
with $\nu_{B} = eB/2 \pi m_{e}c$.
The distribution $f(p)$ and $q$ are found by equating a selected observational 
frequency $\nu$ to the critical synchrotron frequency of emission, $\nu_{c}$, 
to identify a momentum $p = m_{e}c~[2\nu_{c}/(3\nu_{B_{\bot}})]^{(1/2)}$. 
This calculation is performed in the rest frame of the simulated radio galaxy,
with the appropriate redshift correction made for the observation frame.  
It also explicitly takes into account variations in the angle of the magnetic 
field projected on the plane of the sky, $\theta$.  For these simulations all 
flow speeds are subrelativistic, so kinematic Doppler factors and light-travel
times are ignored. Note that because we are explicitly calculating the 
momentum distribution of nonthermal electrons in the simulations, we obtain a 
local slope to the momentum distribution, $q$, by interpolating between
momentum bin centers.  
The synchrotron spectral index $\alpha = (q-3)/2$ obtained from these 
synthetic observations is therefore self-consistent.  
Previously, spectral indices in synthetic 
observations from purely MHD simulations had to be included in an ad hoc 
fashion \citep{Matthews90,Clarke93}.  Surface brightness maps for the 
optically thin emission are produced from our emissivity distributions 
via raytracing through the computational grid to perform line-of-sight 
integrations, thereby projecting the source on the plane of the sky at any
arbitrary orientation.  This method also enables us to compute Stokes 
parameters for the synchrotron emission \citep{Clarke89}, 
as well as the correction for Faraday rotation through the source, making 
detailed polarimetric studies possible.  We have also produced X-ray surface 
brightness maps in the same fashion, by calculating the inverse-Compton (IC) 
emissivity from the interaction between the cosmic microwave background 
radiation and the nonthermal electrons \citep*{Tregillis01}.  
 
	The synthetic observations can be imported into any standard image 
analysis package and subsequently analyzed like real observations.  The 
analysis here was performed using both the \textsc{MIRIAD} and \textsc{KARMA} 
\citep{Gooch95} packages.  For example, it is a straightforward matter to 
construct spectral-index maps from a set of observations over a range of 
frequencies.  To make this exercise as realistic as possible we place the 
simulated object at an appropriate luminosity distance, set to $100$ Mpc 
for the observations included here, although that choice has no influence
on our conclusions.  Because our primary interest here is in identifying 
general trends, the observations are presented at their full resolution with 
very high dynamic range, although it is straightforward to convolve the images
down to lower resolution before making comparisons to true observations.
We note that line-of-sight integrations generally tend to enhance regions of 
flatter emission over steeper emission, and suppress regions of weaker 
emission.  Therefore when investigating our three-dimensional sources we have 
been careful to study multiple orientations in order to identify accidents of 
projection along the line of sight.

  The radio luminosity of our
simulated sources is somewhat arbitrary, as it scales with the ratio of 
nonthermal to thermal particle number densities in the jet, $\delta$. 
This is 
essentially a free parameter in the
simulations as long as the energy density in nonthermal electrons remains
dynamically unimportant.  
As a consistency check, we compute 
the spectral luminosity $L_{\nu}$ at $1.4$ GHz
and compare this value to the jet kinetic luminosity, $L_{j}$.  
In all cases we 
find $L_{\nu}$ to be significantly less than the kinetic luminosity.  For
the time corresponding to that in the images ($t = 4.0$), we find
$L_{\nu}/L_{j} = 4.9 \times 10^{-10} \times (10^{4}~\delta)$ for model 1,
$3.7 \times 10^{-10} \times (10^{8}~\delta)$ for model 2, and 
$4.0 \times 10^{-7} \times (10^{4}~\delta)$ for model 3.  (Recall
$\delta = 10^{-4}$ in models 1 and 3, and $\delta=10^{-8}$ in model 2.)
We find the highest ratio in the model with the strongest
magnetic fields (model 3), and the lowest ratio in the model
with the lowest cosmic ray number densities (model 2).  Note that these 
are the models expected to have the highest
and lowest radio luminosities, respectively.  
The jets are still too dynamically young by the end of the simulations
to have undergone any significant luminosity evolution.  

\subsubsection{``Adiabatic'' Models 1 and 2} \label{adiab.4.3.1.s}

\begin{figure}
\begin{center}
\includegraphics{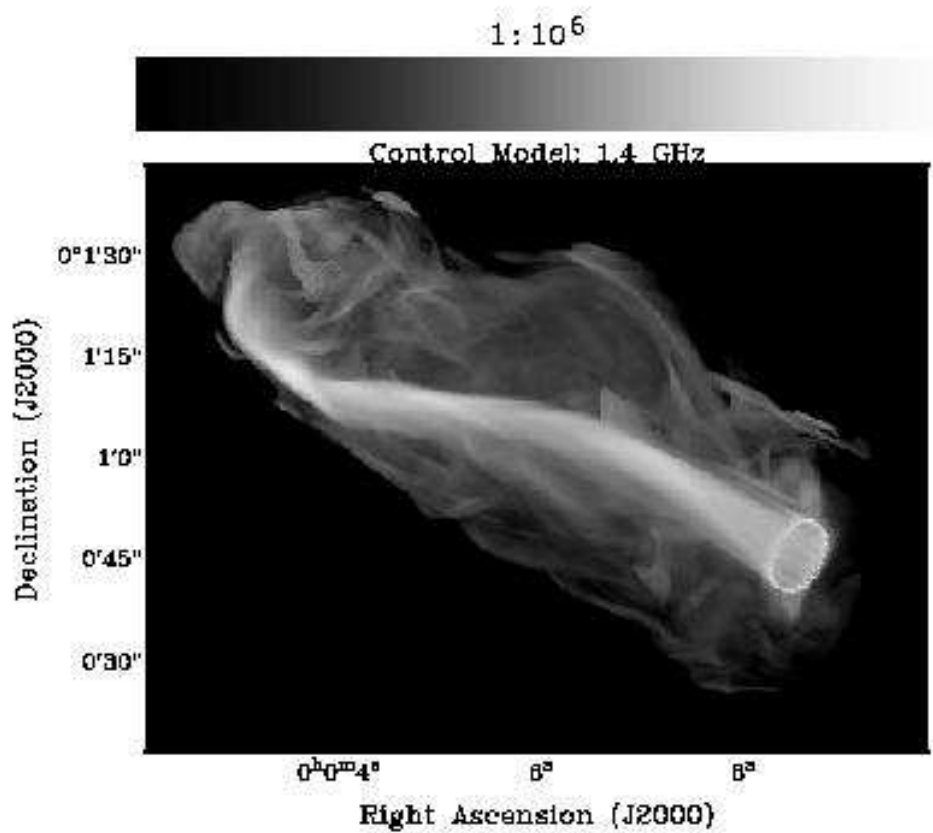}
\figcaption[f4-2.eps]
{Model 1 synchrotron surface brightness map calculated at 
$\nu = 1.4$ GHz, time  $t = 4.0$.
The source orientation in this image corresponds to that in the preceding
images.
\label{model1.f}}
\end{center}
\end{figure}

\begin{figure}
\begin{center}
\includegraphics[scale=0.9]{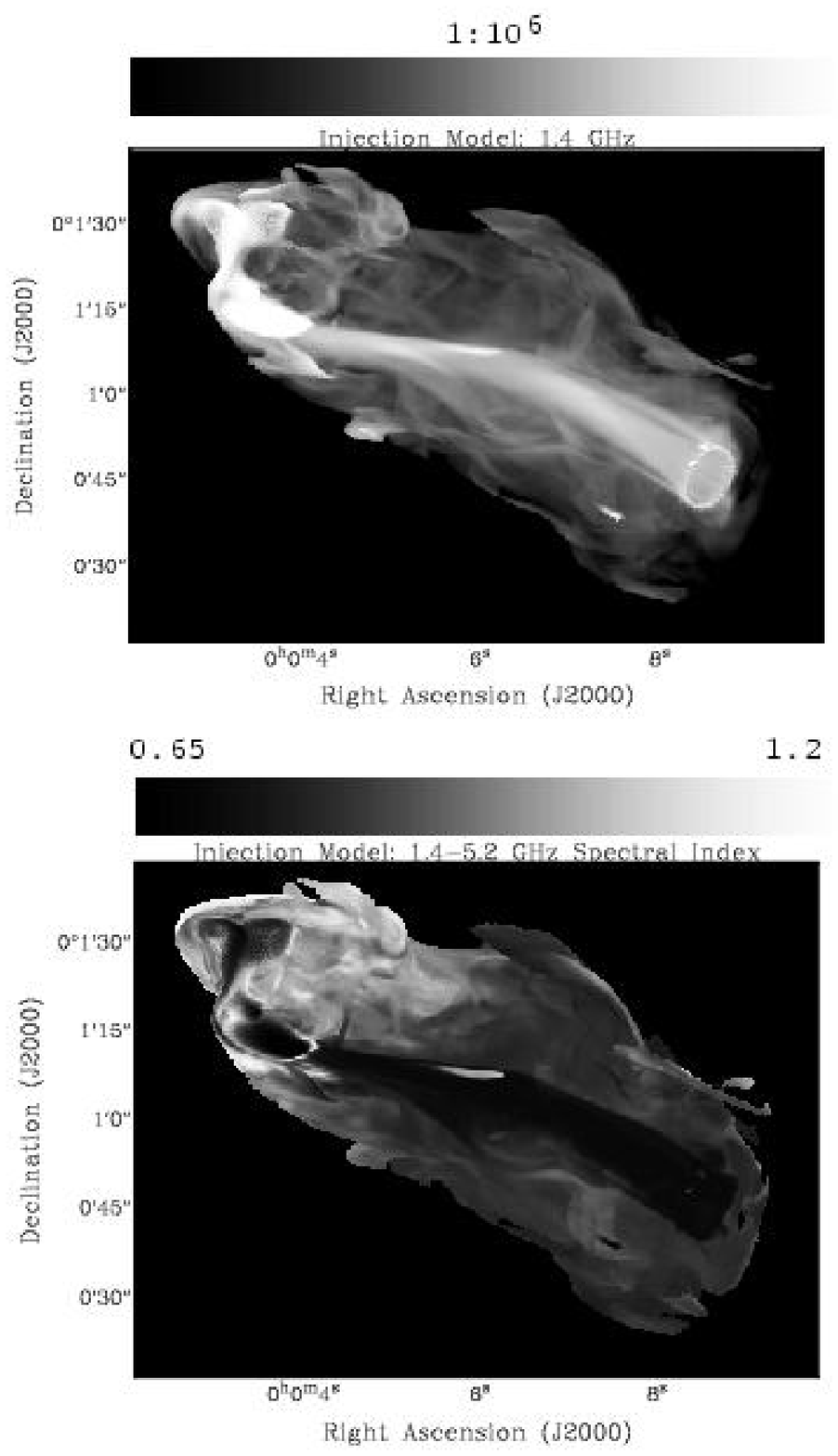}
\figcaption[f5-2.eps]
{(a) Top: Model 2 synchrotron surface brightness map calculated at 
$\nu = 1.4$ GHz, time $t = 4.0$.
(b) Bottom: Model 2 synchrotron spectral index map calculated from surface 
brightness maps at $\nu = 5.2$ GHz and $\nu = 1.4$ GHz, time $t = 4.0$.
\label{model2.f}}
\end{center}
\end{figure}

	Figure \ref{model1.f} shows an image of synthetic 
synchrotron surface brightness at 1.4 GHz for $t=4.0$ in model 1.
Figures \ref{model2.f}a and b show the corresponding image for model 2 and
the two-point spectral index computed from
maps at 1.4 GHz and 5.2 GHz for $t=4.0$ in model 2, respectively.
The orientation of 
the source in these images corresponds approximately to that of the 
computational volume in the volume rendered images. In both cases we have 
restricted the calculation of synchrotron emissivity only to regions where 
the jet color tracer $C_{j} \geqslant 0.99$.  Since $C_{j} = 0$ in the ambient
medium, this selects emission originating only from material that entered
the computational volume by way of the jet.

	The jet and its cocoon are both readily visible in each of the 
surface-brightness images.  The bright ellipse at the lower-right is the 
orifice where the jet enters the computational grid.  Faint brightness 
enhancements are 
visible in the radio jets of both models, the smallest of these also being the 
most prominent.  We find that the enhancements are caused by the very slight 
increase in the jet magnetic field at the oblique shocks, such as the conical
oblique shock which is faintly visible just downstream of the jet orifice in
figures \ref{model1.f} and \ref{model2.f}a.  Another such enhancement can
be made out roughly halfway along the jet in these figures, along the segment
of the jet oriented nearly horizontally with respect to the page.  
It is reasonable to ask if these brightness variations are 
augmented by enhanced particle populations downstream of oblique shocks in the
injection model (recall that injection of fresh particles was allowed to take 
place in model 2).  Oblique shocks with compression ratio $r < 3.14$ would 
inject particles with $q > q_{jet} = 4.4$ into the jet flow.  Only
one tiny enhancement has a corresponding feature in the spectral-index map,  
so we conclude that acceleration at these shocks does not contribute a 
significant enough electron population to have much influence on the 
synchrotron emission.   

	Away from these enhancements the jet itself is of comparable surface 
brightness to the cocoon in model 2, yet the jet clearly dominates over the 
cocoon emission in model 1.  In the latter model, the entire particle 
population enters the grid via the jet, making the jet a relatively abundant 
source of emitting particles.  On the other hand, the initial jet particle 
population in model 2 was much smaller, so the majority of emitting particles 
were injected freshly at shocks.  Relative jet brightness is also artificially
enhanced in all of our models by the dimensions of our computational box.  
Here, the line-of-sight length ratio between jet and lobe material is a factor
of $10$ at best, whereas in real radio galaxies this ratio is likely much 
larger.  

	The jet is only slightly brighter in the head region of model 1 than
it is elsewhere, and comparable to the brightness near oblique shocks within 
the jet.  There does appear to be a hotspot at the jet terminus, but it is
not very prominent.  In fact there is a fairly strong terminal shock at the 
end of the jet at this time, but it is not apparent in the synthetic image. 
We have found that radio hotspots in synthetic observations of these 
simulations do not always correspond to the spatial location of the 
termination shock and vice-versa \citep*{Tregillis01}. The brightness 
increase resulting from compression of the magnetic field and the local 
electron population at the terminal shock does not outstrip that 
at oblique shocks.  Thus from the viewpoint of the observed emissions, there 
is little practical difference between the structure at the end of the jet and
the oblique shocks further upstream at this time in the simulation.

	Such is not the case in the injection model.  The hotspot complex 
in model 2 consists of a compact, bright region 
apparently associated with the termination of the jet, connected by a thin 
bridge of emission to a larger region of enhanced brightness.  We will follow 
convention and refer to the smaller, brighter region as the ``primary'' 
hotspot, and the larger, weaker region as the ``secondary'' hotspot.  The
advantages of combining synthetic observations with the known dynamical history
of a simulated source become apparent as we investigate the nature of these
hotspots.

	Notice 
how the secondary hotspot is elongated in a direction roughly parallel to the 
outer cocoon wall, and normal to the line leading to the primary.  This very 
nicely fits the description of a ``splatter spot'' 
\citep[\eg][]{Williams85,Cox91}.  
In the splatter spot scenario, there is an outflow from the primary hotspot 
that impinges upon the opposite wall of the cocoon, providing an energy 
supply for a secondary hotspot.  The notion of a primary outflow
appears to be an apt description of the 
flow dynamics inferred solely from this single synthetic observation.  It 
leads us to interpret the bridge as a flow which 
connects the primary and secondary hotspots.  However, since there are other
possibilities (for instance, the apparent relationship between the primary
and secondary may only be an accident of projection) it is difficult to know
if this interpretation is correct without further information.  
  
	More direct evidence that the secondary hotspot in figure 
\ref{model2.f}a is the result of an outflow from the primary hotspot
is obtained by considering the jet dynamics.  Analysis 
reveals that the jet is  undergoing a ``breaking'' event as 
described above in section 4.1.  The highest velocity core of the jet is 
severely bent, and the primary hotspot here is in fact associated with the 
shock at the ``elbow''.   The synchrotron emissivity jumps by over four 
orders of magnitude across this shock, going from the faint jet to the bright 
hotspot.  Close inspection of the behavior of the magnetic pressure in the 
same region reveals an increase only by a factor of $\approx 2.6$.  On the 
other hand, injection of fresh energetic particles from the thermal plasma 
passing through this shock increases the population of particles in the 
momentum bin of interest by a factor of $\approx 1.1 \times 10^{4}$ since
a very minimal population was present in the jet at the beginning.  Thus, 
the brightness of the primary hotspot is almost entirely the result of an 
enhanced particle population at the shock and not a dramatic increase in 
magnetic field strength.  The newly-injected electrons flow through the 
bridge into the secondary hotspot, where the magnetic pressure increase over 
the jet value is still $\lesssim 3$.  This outflow from the primary hotspot
is key to the appearance of the secondary hotspot.

	We note here that multiple hotspots are not uncommon in synthetic
observations of the injection model, yet not all secondary hotspots are
splatter spots.  Another model for secondary hotspots is the so-called
``dentist's drill'' model \citep{Scheuer82}.  In this case the jet impinges
upon the cocoon wall upstream of a primary hotspot, forming a new primary
hotspot while the old primary subsequently appears as a secondary hotspot.
This is distinguished from a splatter spot in that the secondary is not
powered by an outflow directly from the primary.  Such a situation occurs
at t=3.6 in the simulation.  Lacking any power supply at all, such a 
disconnected hotspot might be expected to dissipate on a timescale on the 
order of its adiabatic expansion time \citep{Valtaoja84,Lonsdale86}.  Our 
findings support those of \citet{Cox91}, who find that such secondaries may 
last considerably longer than this.  Rather than being downstream of an 
outflow directly from the primary hotspot, they may be powered by material 
that was downstream of the ``elbow'' during a jet ``breaking'' event as 
described in section 4.1.  Dissipation occurs not on the adiabatic expansion
timescale but on the order of the time it takes the disconnected material
near the jet head to reach the secondary.  

	We find  that the advance speed of the terminal shock (when such can 
be easily identified) varies by a factor of several around the mean velocity 
determined by the total time required to traverse the grid.  As mentioned
above, there is not always a one-to-one correspondence between hotspots
and terminal shocks in these synthetic observations.  Nonetheless 
the highest-velocity fluctuations 
correspond to the ``breaking'' events described above when the newly realigned
jet head surges forward.  Often the jet has a strong, transient terminal 
shock and a bright hotspot during these events.

	The jet in these models has a spectral index of $\alpha \approx 0.7$, 
as expected for $q_{jet} = 4.4$.  Both models 
also show spectra flatter than $0.7$ at the primary hotspot, which again is 
as expected if the hotspots are identified as the location of a strong shock.
Yet the absence of radiative cooling in these adiabatic models leads to 
cocoon spectral index distributions that at first glance may seem to be at 
odds with the paradigm for radio galaxy spectra. 

	For model 1, the cocoon spectral index is extremely uniform, and shows
only very minor variations ($\Delta\alpha \approx 0.01$) from the jet value, 
with the cocoon only marginally steeper than the hotspot.  
As described in section 4.2.1, only small amounts of material with
$q < 4.4$ (corresponding to $\alpha < 0.7$) enter the cocoon in model 1.
All of the material in the cocoon is subjected to adiabatic expansion.
The effect is to shift the observed spectrum by equal amounts in $\log{I}$ and
$\log{\nu}$ \citep{Katz-Stone97}.  For a pure power-law spectrum this
introduces no curvature when observing a source at a fixed frequency.
This is in contrast to what was found in the analogous model 1 
in JRE99.  There much of the emission outside of the jet, including the
brightest parts of the cocoon, was associated with spectral indices flatter 
than the $\alpha = 0.7$ within the jet itself.  Because the nonthermal particle
spectra are essentially still power laws, the flow of material into the
adiabatically-expanded magnetic voids does not lead to increased spectral 
curvature in the observed emissions.
	
	For model 2, we find the steepest spectrum material near 
the head of the cocoon, and the spectral index of the cocoon actually flattens 
towards the jet origin, which is completely at odds with what would be
expected based on the standard paradigm for radio galaxy aging.
We found in section 4.2.1 that the weaker but extensive elements of the shock
web inject large populations of steep-spectrum electrons into the cocoon
in model 2.  Just downstream of these weak shocks, emission from
these freshly-injected steep populations can dominate over the unmodified jet 
material because the population transported down the jet is small in this case.
This leads to the appearance of flat-spectrum hotspots as islands in a sea of 
much steeper emission.  The secondary is slightly steeper than the primary 
because of adiabatic expansion.  

	Two effects cause the lobe radio emission in this model to become 
flatter as 
material moves out of the head region.  As the newly-injected steep-spectrum 
material joins the backflow, it mixes with older, flatter material (that 
has either been reaccelerated or injected at a previously existing strong 
shock or has passed directly from the flatter jet into the backflow) 
causing the overall cocoon emission to quickly flatten out as it moves back 
toward the core.  Secondly, as has been mentioned above, the magnetic field 
in the cocoon is generally weak owing to adiabatic expansion into the lobe.
Lower magnetic field values select higher-energy electrons at a given 
observation frequency, which typically results in steeper emission for convex 
spectra, such as those obtained from radiative aging of an initially power-law 
distribution.  However, analysis of color-color diagrams 
\citep[as per][]{Katz-Stone97} 
of these data reveals the cocoon spectra to be concave in the 
injection-dominated case of model 2, as expected. (Recall the absence of
radiative cooling in this simulation.)  Thus, the higher-energy 
electron populations selected by the diminished lobe fields are actually 
flatter than the lower-energy populations.  This also contributes to the 
flattening of the lobe spectrum as material moves out of the head region.  

	It is important to point out here that the presence of both steep-
and flat-spectrum populations along the same line of sight can have a dramatic
effect upon the resulting spectral-index maps.  For a given magnetic field
and nonthermal particle number density, flatter-spectrum populations are more 
likely to dominate the emission over steeper populations, especially
at higher frequencies.  Nevertheless, while the steep emission around the jet 
head varies in prominence as the orientation of the source on the sky is 
varied, the general trend does not disappear.  Rather than being an artifact
of orientation, this trend is a property of the model throughout the length 
of the simulation after the disruption of the Mach disk.

	This differs from what was seen in the analogous axisymmetric model
in JRE99.  There, the general trend was for steepening away from the jet head,
in accord with the usual expectations for radio galaxy spectra.  In that case,
the enforced symmetry prevents the creation of a shock web complex capable of 
generating the prodigious amounts of steep-spectrum material needed to 
dominate emission over the flatter jet material.  The shock web is simply a 
consequence of broken symmetry in the three-dimensional calculation, and
the ``backwards'' spectral index gradient is simply a consequence of the 
injection-dominated particle transport taking place within the shocks of
modest strengths within the web.  This model represents an extreme idealized 
case where injection of fresh particles at shocks dominates the nonthermal 
particle transport, and as such is not meant to represent a real-world radio 
galaxy.  (However \citet{Treichel01} have found that spectral analysis of 
the FRII radio galaxy 3C 438 reveals steeper emission near the head than 
closer to the core.)
A real radio source characterized by these dynamical and transport effects 
would exhibit this kind of spectral index gradient.  The existence of such 
clear features in one model does bring out the diagnostic potentials for 
exploiting the relative importance of local electron injection in weak shocks 
associated with backflows.

\subsubsection{``Radiatively Aged'' Model 3} \label{radage.4.3.2.s}

\begin{figure}
\begin{center}
\includegraphics[scale=0.9]{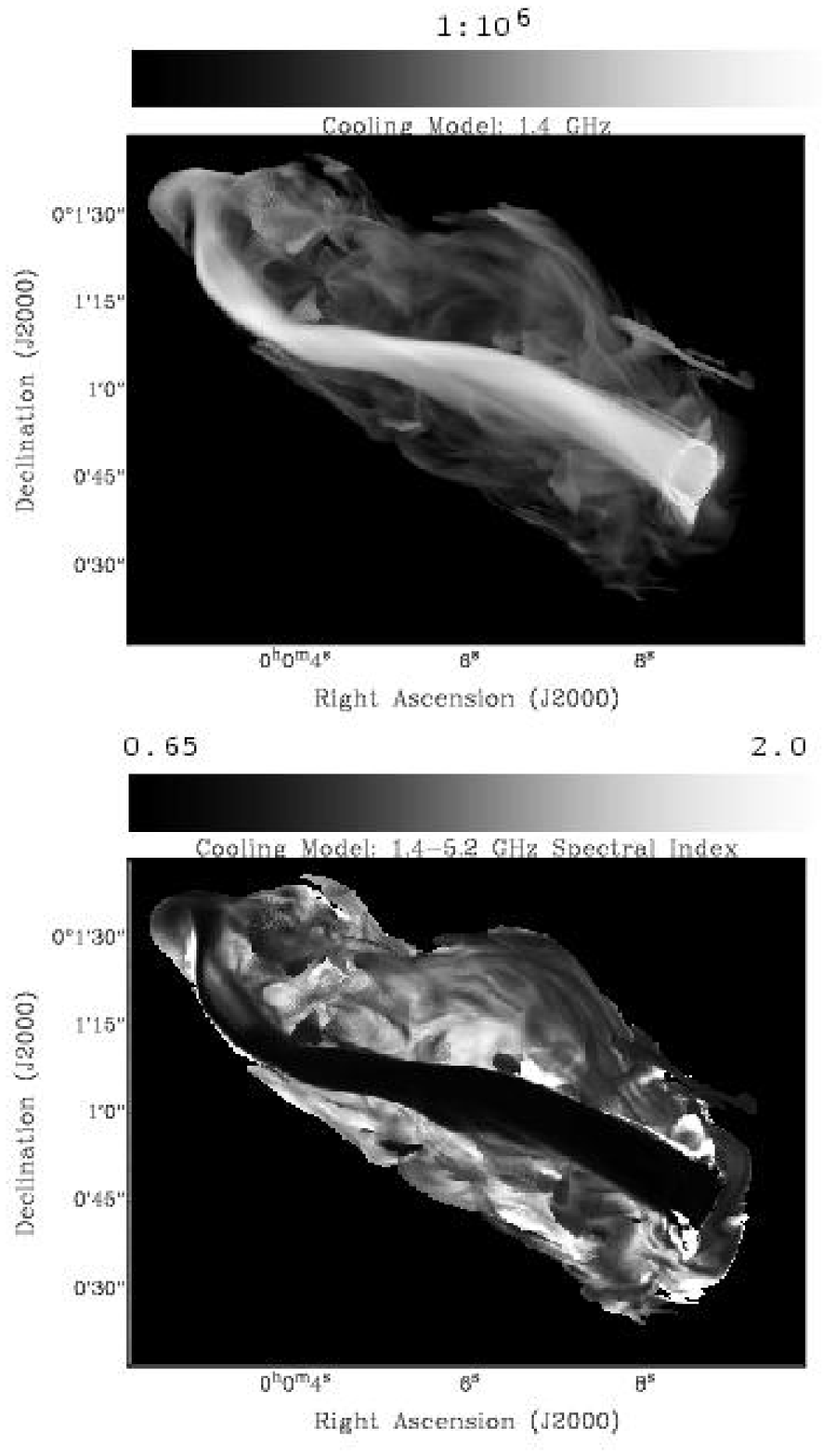}
\figcaption[f6-2.eps]
{(a) Top: Model 3 synchrotron surface brightness map calculated at 
$\nu = 1.4$ GHz, time $t = 4.0$.
(b) Bottom: Model 3 synchrotron spectral index map calculated from surface 
brightness maps at $\nu = 5.2$ GHz and $\nu = 1.4$ GHz, time $t = 4.0$.
\label{model3.f}}
\end{center}
\end{figure}

	Figures \ref{model3.f}a and \ref{model3.f}b show images of synthetic 
surface brightness at 1.4 GHz and the synchrotron spectral index computed from
maps at 1.4 GHz and 5.2 GHz for $t = 4.0$ in model 3.  The source orientation 
in these figures is identical to that in figures \ref{model1.f} and 
\ref{model2.f}.  As above we have restricted the emissivity calculation to 
jet material only ($C_{j} \geqslant 0.99$), but we have also restricted the 
calculation to zones where $q \leqslant 8$ ($\alpha \leqslant 2.5$). This 
condition is a reasonable way to avoid occasional spurious numerical effects arising close to electron cutoffs, since extremely steep populations would 
likely be weak emitters anyway, and results in the exclusion of only a small 
number of zones.
	
	Again we find the jet core dominates the source for the same reasons 
as explained in section 4.3.1.  Since in this model new energetic particles 
are not being introduced into the flow at shocks, the brightness variations 
are largely tracers of magnetic field enhancement, as in model 1.  Thus we see
the familiar pattern of brightness increases near oblique shocks in the 
jet, without corresponding features in the spectral index map.  There 
is also a dramatic variation of surface brightness throughout the cocoon, 
where high-surface-brightness filaments wind through fainter material.

	The key effect that differentiates between brightness maps of
models 1 and 3 is the mixing along lines of sight between flatter and 
steeper (radiatively aged) populations.  Since this mixing will tend to 
emphasize the former, we expect to see more similarities between the 
surface brightness maps of models 1 and 3 beyond those expected on the basis
of the identical dynamics in those two models.  And indeed, 
while the jet and cocoon are both readily visible as in models 1 and 2,
the jet terminal structure is much more akin to that of model 1 than the
hotspot complex of model 2.  In fact the jet terminus associated with 
the primary hotspot in model 2 exhibits only a very minor brightness 
enhancement beyond the overall brightness of the jet, and the region 
corresponding to the secondary hotspot is much diminished in intensity.  Again
we emphasize the fact that these simulations are dynamically identical, and 
thus the jet terminus does host a strong terminal shock at this time in model 
3 as well.  A tiny ``sliver'' of material can be seen flowing out of the 
terminal shock region, along the lower edge of the contact discontinuity, 
without suffering even the mild brightness enhancement associated with the 
shock in this model.  This flow element is not prominent in model 2 as it 
represents material that passed from the jet into the cocoon without being 
shocked, but it is also visible in model 1.

	While the surface brightness distribution is similar to that of the
control model, the spectral index map differs substantially.
The jet itself has aged noticeably, steepening from $\alpha_{jet} = 0.7$ to 
$\alpha \approx 0.8$ just prior to the primary hotspot.  The jet does appear 
to experience appropriately stronger cooling in the stronger magnetic field 
near the primary hotspot, with $\alpha \gtrsim 1$ there.  Virtually all of the
cocoon material is steeper yet, with $\alpha \approx 1.0 - 2.0$.  The 
distribution of indices strongly suggests almost random mixing between the 
steepest and flattest regions rather than a clear gradient away from the jet 
head as in model 2.  In fact, close inspection of the spectral index map near 
the primary hotspot clearly shows the same tiny sliver of jet material 
mentioned above, which flows past the primary into the secondary hotspot 
without suffering the aggravated cooling experienced by most electrons in this 
region.  Again we see the result of highly complex flow patterns in and around
the jet head, this time allowing small amounts of material to escape aging in 
the high field regions associated with a hotspot, thereby injecting 
comparatively ``young'' material into the cocoon.  Strong mixing between this
material and material that has been processed through the high-field regions 
of previous hotspots and filaments in the backflow is the source of this 
complex spectral index pattern.  

	Also, these simulated sources are fairly young in terms of their 
dynamical histories, so individual events (\eg breaking of the jet) in their 
evolution can still have strong effects on the overall appearance of the 
source.  The moderately steep region in the upper-center part of the spectral 
index map (above the jet) for this model is the result of a single, brief 
and strong magnetic
field enhancement owing to shearing at the jet end at an earlier time in the 
source evolution.  As explained in section 4.2.3, the source dynamics leads 
to sporadic creation of strong-cooling regions and large magnetic ``voids''.
The lower fields of the voids lead to emission from higher-energy particles
(for a given observation frequency), and thus emphasize the cooling-induced
spectral curvature in this model.  If the flows conspire to keep large 
numbers of particles in these volumes for significant lengths of time, the 
overall aging rate could be much slower than would be estimated based on 
regions of high emissivity, where the fields are often stronger than
average.

  	Unlike the previous models 1 and 2, the hotspots here in model 3 do 
not correspond to significantly flatter spectra.  As explained in section 
4.2.3, radiative aging of material transported down the jet and the lack of 
fresh particle injection at the relatively rare strong shocks are contributing
factors.  Here synchrotron cooling in the moderately enhanced magnetic field 
of the hotspot has overcome reacceleration at the cospatial strong shock.  
On top of mixing this exacerbates the difficulty in finding a spectral 
gradient away from the jet head region.  This contrasts with what might 
have been expected based on the analogous axisymmetric model (JRE99), where 
the terminal shock and indeed the oblique shocks within the jet flattened the 
jet material, even in their strong-cooling simulation.  This is possible in 
the axisymmetric model because there as here the jet spectrum undergoes some
radiative aging between the orifice and the terminal shock, but there the
oblique shocks are just strong enough to reaccelerate the slightly-aged
jet material.  While still highly variable the 
terminal shock in the axisymmetric simulations is generally stronger than 
those in the shock web complex here.  

	In model 1 there were hardly any spectral index variations
in the cocoon to correlate with brightness variations.  Here, however,
there is in general a nice correlation between brightness and spectral
index, in the sense that brighter material is typically flatter than 
surrounding fainter material.  This relationship is not so clear-cut in model 
2, where the hotspots are certainly flat, yet the prevalence of steep-spectrum
emission from the region around the jet head resulting from substantial 
electron injection at weak shocks does not appear to lead to a 
diminished brightness compared to other parts of the cocoon.  

\subsubsection{General Comments: Electron Transport and RG Dynamics\\
Application to Real Radio Galaxies}\label{gcom.4.3.3.s}

	In each of the models considered in the previous sections the cocoon 
appears to have an intricate network of bright filamentary structures spread 
throughout material of lower surface brightness.  The magnetic intermittency 
($\frac{<B^{4}>}{<B^{2}>^{2}}$)
calculated for the entire cocoon volume at this time is approximately $100$, 
indicating that the magnetic field structure is also extremely filamentary.
Most high magnetic field regions outside of the jet core are confined to a 
relatively small volume of flux tubes that wind through the source.  
Application of ``spectral tomography'' techniques \citep{Katz-Stone97} to 
these data reveal no readily-apparent overall structure in the cocoon 
filaments, although there is a wealth of fine structure in both surface 
brightness and the spatial distribution of various spectral indices.

	In 1989, Scheuer put forth some simple but compelling arguments that
adiabatic compression of the magnetic field alone could increase hotspot
brightnesses by even 2 orders of magnitude over the jet, and that in fact
it is almost surprising that jets can be seen at all.  Model 1 presented here 
is most similar to the situation considered for these arguments. Even there, 
however, we see a jet very much brighter than the cocoon, roughly ten times so
at the time shown in the figures.  We find that the various shocks in the 
complex near the jet head rarely compress the field to such a great extent, 
even when they are strong.  Rather, shearing of the flow near the end of the 
severely-bent jet is the leading creator of the filamentary strong-field 
regions in the backflow.  As has been pointed out above, in these models 
adiabatic expansion into the lobe acts to reduce the brightness of the cocoon. 
The issue here is the question of field compression in the terminal shock, 
which is often weak and hard to identify.  This reminds us that the working 
surface in these models is not just the terminal shock. 

	By restricting the emissivity calculations here to the volume inside
the contact discontinuity (where $C_{j} = 1$), we have ignored emission arising
from shocked ambient medium.  Yet it is interesting to note that even when the
effective area of the working surface of the jet head is augmented by the 
time-averaged effects of precession, the thermal plasma flux through the bow 
shock in the ambient medium may still be larger than the analogous flux 
through other shocks in the system.  
Significant nonthermal emission might therefore arise from shocked ambient 
medium between the jet contact discontinuity and the bow shock if the magnetic
field there is strong enough.  Particularly 
in our injection-dominated model 2, we see a considerable amount of emission 
in the ambient medium, when it is not excluded.  This emission evidences 
variations along the length of the source consistent with irregular cycles of 
weakening and strengthening of the bow shock.  Naturally this emission 
depends strongly on the character of the ambient magnetic field, and it
is difficult to know in detail how a different treatment of this field in our 
models would alter the effect.  Note, however, that there is mounting evidence
that magnetic fields in some clusters may approach 
$10 \mu $G \citep[\eg][]{Clarke00}, yet in 
at least one very powerful source, Cygnus A, the bow shock has been reported 
to be radio quiet \citep{Carilli88}.  If this is generally true, then we may 
in the future be able to place constraints on conditions in the ambient medium
and properties of the jet plasma.  Doing so would require a more realistic 
model for the magnetic field in the ambient medium as well as in the jet 
itself.

\section{CONCLUSIONS} \label{conc.s}

	We have applied the nonthermal particle transport scheme developed by 
JRE99 to the study of three-dimensional MHD flows, in an attempt to 
gain insights into the way that the large-scale dynamical processes in radio 
galaxies affect the relativistic electron populations that reside within.  We 
have examined the role of adiabatic cooling, radiative cooling, and first-order
Fermi acceleration and injection of fresh particles at shocks.  In order to
illustrate better the connection between the flow dynamics and the relativistic
electrons, we have introduced a set of synthetic observations that for the 
first time compute an approximate synchrotron emissivity directly from the 
local magnetic field and nonthermal particle distributions.  This work is an 
extension of the axisymmetric studies cited above.  The most important and 
striking new findings are listed below.

	1. The spatial and temporal shock structure associated with the jet 
head in these studies is extraordinarily complex.  The notion of a single, 
simple, strong terminal shock is applicable only rarely after the flow has 
begun to assert its true multidimensional nature.  This has profound 
consequences for the electron populations processed through these shocks and 
also for the associated synchrotron radiation.  This cautionary note applies 
not only to sources that display strong evidence for precession, but virtually
any sources with complex morphologies.  Our precessed jets show evidence for 
episodic reappearance of a transient strong terminal shock and subsequent 
sudden advances of the jet head.  The complexity of driven multidimensional 
flows \citep{Sato96} makes similarly complicated effects in other scenarios 
quite likely.

	2. This so-called ``shock-web complex'' may lead to spectral 
distributions that confound interpretation in terms of the standard model for 
radio galaxy ``spectral'' aging.  In the case of injection, this may occur by 
way of introducing a significant population of steep-spectrum particles on 
timescales that would not admit such steepening via radiative processes. Even 
in situations where radiative cooling effects are not negligible, a  
shock web may complicate interpretations by allowing radiatively-aged 
electron populations to flow from the jet directly into the backflow without 
getting reaccelerated at a shock, even inside a hotspot; conversely, it may 
allow jet material to escape from a hotspot without being subjected to the 
strong magnetic field there.  All of this underlines the fact that the 
emissions we observe from real radio galaxies yield information about not just
the current state of the nonthermal particles, but also about their dynamical 
histories.

	3. The cocoons of these sources are threaded by strong-magnetic-field 
filaments, but most of the volume is occupied by field values that are 
greatly diminished below the nominal values in the jet.  While compression 
and shearing can enhance the local magnetic
field, adiabatic expansion into the lobe with subsequent reduction in field 
strength appears to be the dominant effect.  Such large, low-field volumes
further tend to confuse spectral-aging analyses by extending the nominal
radiative lifetime of nonthermal particles beyond what would be inferred based
on regions of high emissivity, which mostly represent places where the
fields are strongest.

	4. Overall, the elaborate dynamical behaviors in these models
make the detailed histories of the nonthermal particles very difficult to 
capture succinctly in broad generalizations.  We found that even in the
adiabatic control model (model 1), the energetic electrons in the cocoon do 
not share the same history of shock acceleration.  Yet when fresh particle 
injection at shocks dominates the nonthermal transport (model 2), we cannot 
even simplify matters by restricting our attention to only the strongest 
shocks.  Similarly, fitful amplification of the magnetic field leads to 
varying histories of radiative energy loss among energetic electrons in the 
cooling model (model 3).   

	We have restricted our attention here to the some of most general and 
wide-ranging issues, intending to lay the groundwork for further, more 
in-depth examinations.  Later papers will apply the tools and methods shown 
here to more specific and detailed questions regarding the nature of radio 
galaxies.  The insights gained from these idealized, extreme cases for 
nonthermal particle transport will be valuable in studies that generalize the 
transport model and aim to identify physical behaviors seen in specific 
observed objects. This first pass has demonstrated the power and necessity of 
explicitly modeling energetic particle transport effects in bulk flows in any 
attempt to convert dynamics to nonthermal emissions.

\acknowledgements
	The work by T.W.J. and I.L.T. was supported by the NSF under grants 
AST96-16964 and AST00-71176 and by the University of Minnesota Supercomputing
Institute.  The work by D.R. was supported in part by KRF through grant 
KRF-2000-015-DS0046.  We gratefully acknowledge Larry Rudnick for many
helpful comments and discussions.

\clearpage
\begin{deluxetable}{cccccc}
\tablewidth{0pt}
\tablenum{1}
\tablecolumns{6}
\tablecaption{Summary of Simulations\label{param.t}}

\tablehead{  
\colhead{Model\tablenotemark{a}} &
\colhead{ID} &
\colhead{In-flowing} &
\colhead{Shock Injection}  &
\colhead{Cooling} &
\colhead{$B_{x0}$}\\
& & Electrons\tablenotemark{b}~($b_{1}$)& Parameter~($\epsilon$) & 
Time\tablenotemark{c}~($Myr$) & ($\mu G$) \\
}

\startdata
1....... & Control & $10^{-4}$ & $0.0$ & $1.63 \times 10^{4}$ & $0.39$\\ 
2....... & Injection & $10^{-8}$ & $10^{-4}$ & $1.63 \times 10^{4}$ & $0.39$\\
3....... & Cooling & $10^{-4}$ & $0.0$ & $54$ & $5.7$
\enddata

\tablenotetext{a}{All models used external Mach $80$ jets 
($M_{j} = u_{j}/c_{a} = 80$), 
which corresponded to a velocity of $0.05$ c, and a density contrast
$\eta = \rho_{j}/\rho_{a} = 0.01$; thus the internal jet Mach number is $8$.
Units derive from  $r_{j}=1$ 
(representing 2 kpc in physical units), an ambient density, $\rho_{a}=1$, and 
a background sound speed, $c_{a}=(\gamma P_{a}/\rho_{a})^{1/2}=1~(\gamma=5/3)$.
The initial axial magnetic field was $B_{x0}$ ($\beta = P_{a}/P_{b} = 
100$) in the ambient medium.  The jet also carried an additional toroidal 
field component, $B_{\phi} = 2 \times B_{x0}(r/r_{j})$.  The spectrum of 
nonthermal particles in the jet was set to a power law with momentum slope 
$q=4.4$, which corresponds to a synchrotron spectral index $\alpha = 0.7$.  
The nonthermal particle distribution was specified by $N = 8$ momentum bins in 
all three models.}

\tablenotetext{b}{Ratio of nonthermal to thermal electron density in the 
incident jet flow.}

\tablenotetext{c}{Time for electrons to cool below momentum 
$\hat p= 10^{4}m_{e}c$ in the background magnetic field $B_{x0}$.  In these
simulations the time unit $r_{j}/c_{a}$ corresponds in physical units
to approximately $10$ Myr.}

\end{deluxetable}
\clearpage


\clearpage

\end{document}